\documentclass[
reprint,
superscriptaddress,
amsmath,amssymb,
aps,
prd,
floatfix,
]{revtex4-2}

\usepackage{graphicx}
\usepackage{dcolumn}
\usepackage{bm}
\usepackage[english]{babel}
\usepackage[utf8]{inputenc}
\usepackage[T1]{fontenc}
\usepackage{microtype}
\usepackage[table,dvipsnames]{xcolor}
\usepackage[
colorlinks=true,
breaklinks=true,
hyperfootnotes=true,
]{hyperref}
\usepackage{booktabs}
\usepackage{subfigure}
\usepackage{ulem}
\usepackage{multirow}

\begin{document}
	
	\preprint{APS/123-QED}
	
	\title{Compact Binary Systems Waveform Generation with Generative Pre-trained Transformer}
	
	\author{Ruijun Shi}
	\altaffiliation{Equal contribution}
	\affiliation{Department of Astronomy, Beijing Normal University, Beijing 100875, China}
	
	\author{Yue Zhou}
	\altaffiliation{Equal contribution}
	\affiliation{Peng Cheng Laboratory, Shenzhen 518055, China}
	
	\author{Tianyu Zhao}
	\affiliation{Department of Astronomy, Beijing Normal University, Beijing 100875, China}
	
	\author{Zhoujian Cao}
	\affiliation{Department of Astronomy, Beijing Normal University, Beijing 100875, China}
	
	\author{Zhixiang Ren}
	\altaffiliation{Corresponding author}
	\email{renzhx@pcl.ac.cn}
	\affiliation{Peng Cheng Laboratory, Shenzhen 518055, China}
	
	\date{\today}
	
	\begin{abstract}
		Space-based gravitational wave (GW) detection is one of the most anticipated GW detection projects in the next decade, which promises to detect abundant compact binary systems. At present, deep learning methods have not been widely explored for GW waveform generation and extrapolation.
		To solve the data processing difficulty and the increasing waveform complexity caused by the detector's response and second-generation time-delay interferometry (TDI 2.0), an interpretable pre-trained large model named \textbf{CBS-GPT} (\textbf{C}ompact \textbf{B}inary \textbf{S}ystems Waveform Generation with \textbf{G}enerative \textbf{P}re-trained \textbf{T}ransformer) is proposed. 
		For compact binary system waveforms, three models were trained to predict the waveforms of massive black hole binaries (MBHB), extreme mass-ratio inspirals (EMRIs), and galactic binaries (GB), achieving prediction accuracies of at most 99\%, 91\%, and 99\%, respectively.
		The CBS-GPT model exhibits notable generalization and interpretability, with its hidden parameters effectively capturing the intricate information of waveforms, even with the complex instrument response and a wide parameter range.
		Our research demonstrates the potential of large models in the GW realm, opening up new opportunities and guidance for future researches such as complex waveforms generation, gap completion, and deep learning model design for GW science.
	\end{abstract}
	
	\maketitle
	
	\section{Introduction}
	\label{sec:intro}
	\begin{figure*}
		\centering
		\includegraphics[width=1.0\linewidth]{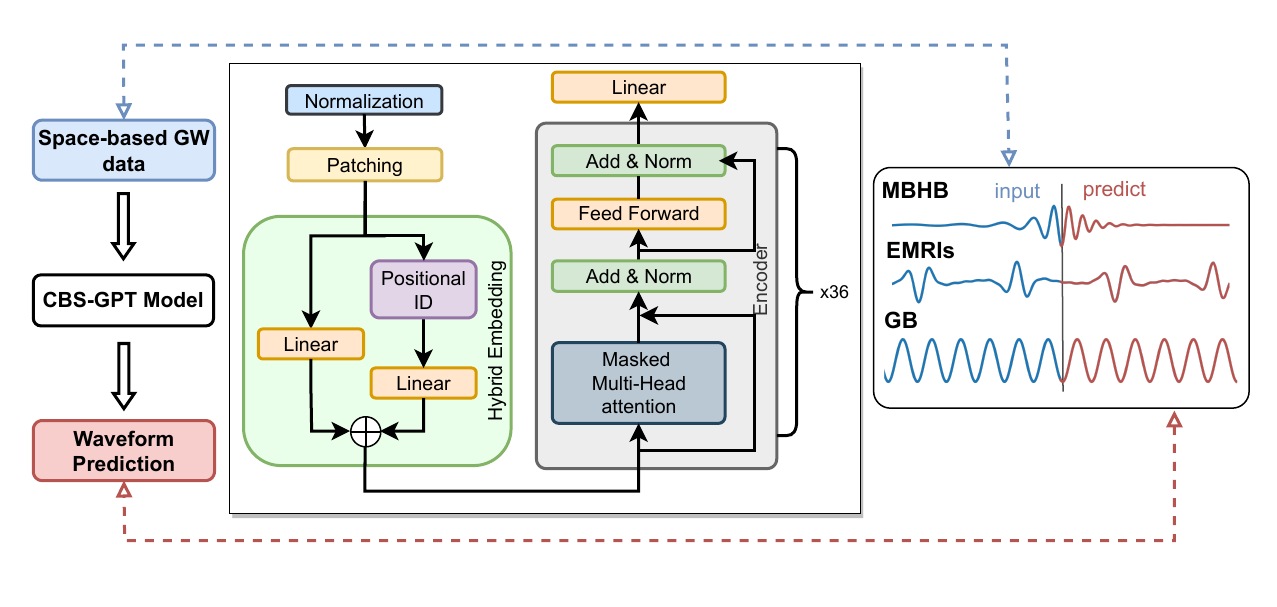}
		\caption{\textbf{Overview of CBS-GPT.} 
            The CBS-GPT model was trained separately for three kinds of GW sources (MBHB, EMRIs, and GB). The subsequent waveform can be extrapolated after feeding its corresponding preceding waveform into CBS-GPT. 
            Details of data and model description are in Section \ref{sec:met}.}
		\label{fig:GPT-model}
	\end{figure*}
	
	The first direct detection of a binary black hole merger (GW150914) \cite{Abbott2016,abbottGW150914FirstResults2016} by the Laser Interferometer Gravitational-Wave Observatory (LIGO) has opened an innovative window to understand the universe, which provides direct evidence for the validity of Einstein's General Relativity. 
	Gravitational wave (GW) observations will clarify many questions in astrophysics, cosmology, and fundamental physics \cite{schutzDeterminingHubbleConstant1986, sathyaprakashPhysicsAstrophysicsCosmology2009, kleinScienceSpacebasedInterferometer2016,  tamaniniScienceSpacebasedInterferometer2016, bartoloScienceSpacebasedInterferometer2016,  capriniScienceSpacebasedInterferometer2016, babakScienceSpacebasedInterferometer2017b, arunNewHorizonsFundamental2022}. 
	So far, the ground-based GW detectors have reported over a hundred compact binary coalesces (CBC) events \cite{TheLIGOScientificCollaboration2021}, and recently Pulsar Timing Array (PTA) has also successfully detected sound evidence of the existence of Stochastic GW Background \cite{Agazie2023, xuSearchingNanoHertzStochastic2023,Antoniadis2023, Reardon2023}. 
	To gain a deeper understanding and an overall picture of GW cosmology \cite{bailes_gravitational-wave_2021}, the field of low-frequency GWs needs to be widely covered. 
    
    The space-based GW detection avoids terrestrial noise \cite{matichard_seismic_2015} and makes the detection of low-frequency ($10^{-4} - 0.1 \mathrm{Hz}$) GW signals more promising. 
    Spaced-based GW detectors like Laser Interferometer Space Antenna (LISA) \cite{amaro-seoaneLaserInterferometerSpace2017}, Taiji \cite{huTaijiProgramSpace2017, renTaijiDataChallenge2023a} and Tianqin \cite{Luo2016} have been planned and are scheduled for the 2030s. 
	In particular, future space-based GW detection is expected to detect a richer variety of GW sources including massive black hole binaries (MBHB), extreme mass-ratio inspirals (EMRIs), and galactic binaries (GB) \cite{amaro-seoaneLaserInterferometerSpace2017}.
 
     GW signals are extremely weak and usually buried in instrumental noise. With the improvement of detector sensitivity and the increasing amount of data, the computational complexity and timeliness demands for detection and parameter estimation are growing, which are challenging problems for traditional methods that based on computing power of central processing unit (CPU).
	With the rapid developing of graphics processing unit (GPU) computing power, Artificial Intelligence (AI) methods have shed some new light on this issue.
	Specifically, AI techniques have been successfully applied in various subjects such as GW signal detection \cite{wangGravitationalwaveSignalRecognition2020, georgeDeepLearningRealtime2018, georgeDeepNeuralNetworks2018,gabbardMatchingMatchedFiltering2018, badgerDictionaryLearningNovel2023, zhangDetectingGravitationalwavesExtreme2022a}, parameter estimation \cite{daxRealTimeGravitationalWave2021a,  gabbardBayesianParameterEstimation2022,wangSamplingPriorKnowledge2022}, signal extraction and noise reduction \cite{ weiGravitationalWaveDenoising2020a, colganEfficientGravitationalwaveGlitch2020, torres-forneDenoisingGravitationalWave2016, akhshiTemplatefreeApproachWaveform2021, Ren2022, zhao_space-based_2023, weiGravitationalWaveDenoising2020, chatterjeeExtractionBinaryBlack2021a} with promising results. 
	Additionally, the target of space GW detectors is also one type of complex and multi-scale waveforms (such as MBHB, EMRIs, and GB). 
	Some previous studies focused on generating  binary black hole (BBH) waveforms. 
	Lee et al. \cite{leeDeepLearningModel2021} employed a Recurrent Neural Network (RNN) that is capable of generating BBH waveforms during the merging and ringdown phases of non-spinning binary black hole coalescence. 
	Khan et al. \cite{khanInterpretableAIForecasting2022} demonstrated that a vanilla transformer can learn quasi-circular and non-precessing BBH waveforms.
	Similarly, Chua et al. \cite{chuaReducedorderModelingArtificial2019} used a greedy algorithm to build a reduced basis, enabling the rapid generation of BBH waveforms. 
	Recently, large-scale language models (LLM) based on attention mechanism have shown their tremendous power in computer vision (CV) and natural language processing (NLP) \cite{devlinBERTPretrainingDeep2019, Brown2020LanguageMA, Liu2021}. Some studies indicate that similar architectures can be applied to the GW data analysis \cite{zhao_space-based_2023, Ren2022}.  
    Space-based GW detectors will observe more signals along with complex difficulties such as source confusion, gaps, and glitches \cite{lisapathfindercollaborationTransientAccelerationEvents2022a}. It is critical to provide a set of data processing tools to address these issues. Deep learning holds promise for meeting these challenges.

    In contrast to previous studies on AI waveform generation, which had limitations in considering the second-generation time-delay interferometry (TDI 2.0) responses, our paper takes a step further.
    Moreover, the parameter range of waveforms in prior investigations was relatively narrow.
	In our paper, we are committed to further investigation on more complex waveforms and train a model to facilitate solving downstream problems. 
	We introduce \textbf{CBS-GPT} (\textbf{C}ompact \textbf{B}inary \textbf{S}ystems Waveform Generation with \textbf{G}enerative \textbf{P}re-trained \textbf{T}ransformer) model, which is an interpretable, transformer-based, and self-supervised large model for prediction of compact binary sources (MBHB, EMRIs, and GB). 
	In CBS-GPT (Figure \ref{fig:GPT-model}), patching and hybrid embedding mechanisms are proposed for full extraction of waveform features. By utilizing the self-attention mechanism and mean square error loss, CBS-GPT is trained for each GW waveform source.
	The experiment results illustrate that CBS-GPT can accurately predict the subsequent waveform based on the input waveform. In this study, two models were trained to achieve extrapolation with different input-to-prediction length ratios.  In the 20:1 extrapolation, the average overlap between the predicted and target waveforms of MBHB, EMRIs, and GB reaches 0.981, 0.912, and 0.991, respectively. In the 1:1 extrapolation, the average overlaps reached 0.990, 0.807, and 0.992 for MBHB, EMRIs, and GB, respectively.
	We have also discovered that waveform complexity can significantly influence the model's prediction performance, and CBS-GPT can match the key frequencies effectively. 
    Finally, through attention map visualization and correlation calculation, we discover that the attention map and its corresponding waveform present similar periodic distribution, which illustrates that CBS-GPT is able to learn waveform features even under the complex instrument response and a wide parameter range.
	
	The rest of this paper is organized as follows. 
	Section~\ref{sec:met} describes data generation and the CBS-GPT model architecture. 
	In Section~\ref{sec:res}, we present our overlap and attention map results, and discuss interpretability outcomes as well as potential applications. 
	Finally, Section~\ref{sec:con} highlights our findings based on the results.
	
	\section{Methodology}
	\label{sec:met}
	
	\subsection{Data}
	\begin{figure*}
		\centering
		\subfigure[MBHB]{
			\label{fig:TDI_case_MBHB}
			\includegraphics[width=0.95\linewidth]{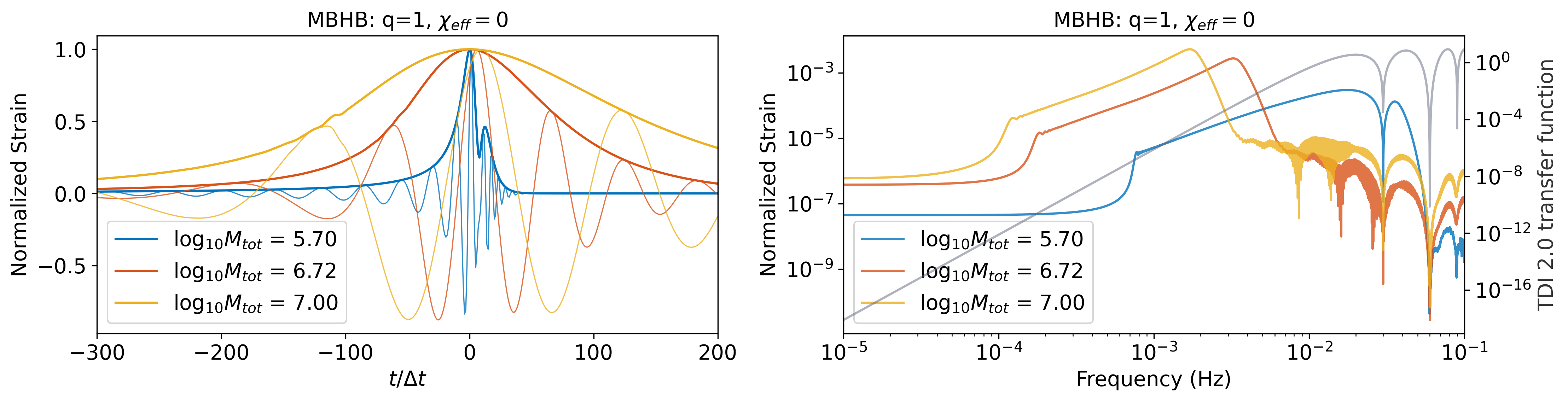}}
		
		\subfigure[EMRIs]{
			\label{fig:TDI_case_EMRI}
			\includegraphics[width=0.95\linewidth]{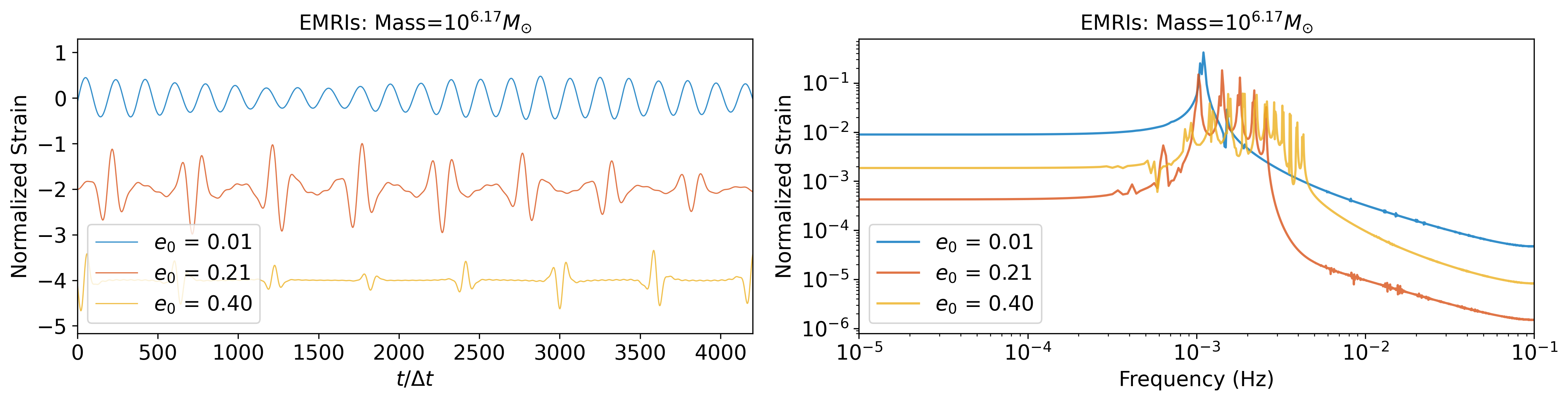}}
		
		\subfigure[GB]{
			\label{fig:TDI_case_GB}
			\includegraphics[width=0.95\linewidth]{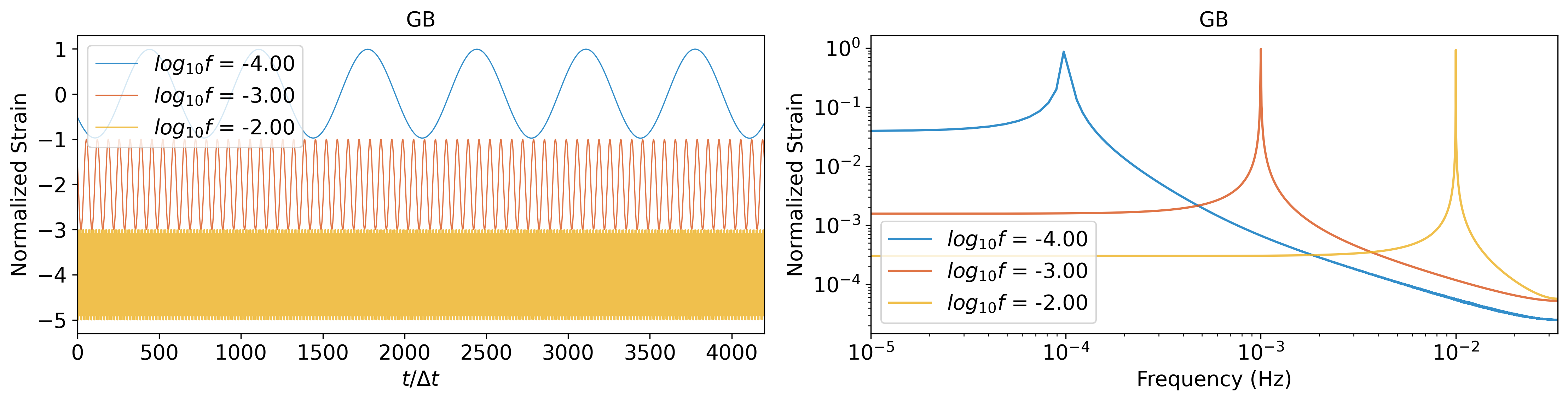}}
		
		\caption{\textbf{TDI2.0 response complicates waveforms.} 
			To simplify waveform comparison here, all waveforms were standardized to a maximum amplitude of 1. The $\Delta t$ in the figure represents the sampling rate. 
			The effects of different parameters on time and frequency domain are shown on the left and right panels.
			(a) \textbf{MBHB waveforms} at different $M_{tot}$. At high frequencies, the TDI response function has a greater impact. The gray line represents the TDI 2.0 transfer function in the frequency domain.
			(b) \textbf{EMRIs waveforms} at different $e_0$. As the eccentricity increases, the EMRIs waveform becomes more and more complex in the frequency domain.
			(c) \textbf{GB waveforms} at different $f$. The GB signal is relatively simple and is a single-frequency signal.
		}
		\label{fig:TDI_complex}
	\end{figure*}
	
	Space-based GW detectors' targets are GW signals at frequencies of $[10^{-4}, 0.1]\text{Hz}$. 
	We focus on three compact binary sources that are of major interest for LISA: MBHB, EMRIs, and GB. Figure \ref{fig:TDI_complex} displays data examples. 
	Detailed information of the data generation process is given below.

\subsubsection{MBHB}
    MBHB are one of the space-based GW detector's main detection targets \cite{amaro-seoaneLaserInterferometerSpace2017}. 
	In this paper, \texttt{SEOBNRv4\_opt} \cite{Bohe2017} ($l=m=2$ mode) is used to generate the MBHB waveforms. 
	The parameter space of the MBHB dataset is shown in Table \ref{tab:MBHB}. 
	In Figure \ref{fig:TDI_case_MBHB}, the TDI 2.0 transfer function significantly affects high-frequency transmissions due to the lower total mass of MBHB. 
	Firstly, we generate MBHB time-series waveforms with a length of 20,000 points with a sampling rate of 5 seconds. 
    We train two models with different input-to-prediction token length ratios. The two scenarios are referred to as 20:1 and 1:1 extrapolation in the subsequent sections. Table \ref{tab:seq_info} summarizes token information for each source. The 20:1 extrapolation involved predicting the subsequent 200 points after merging with an input sequence of the preceding 4000 points. The 1:1 extrapolation predicted the same 200 points after merging, but with an input sequence limited to the preceding 200 points.
	During the inference phase, the 4,000 valid points (or 200 valid points) before the merge time are fed into CBS-GPT to predict the succeeding 200 points, hence achieving a 20:1 extrapolation (or 1:1 extrapolation) prediction of the MBHB waveforms. 
	
	\begin{table*}
		\caption{Parameters distribution of training dataset and test dataset}
		\label{tab:1}
		\centering
		\subtable[Parameter space of MBHB dataset]{
			\label{tab:MBHB}
			\begin{tabular}{p{2.5cm}p{10cm}p{4cm}}
				\toprule
				Parameter    & Description\centering & Parameter distribution \\
				\midrule
				$M_{tot}$   & Total mass of massive black hole binaries $m_1 + m_2$   & log-Uniform $[5.5, 7]M_{\odot}$  \\
				$q$         & Mass ratio $\frac{m_2}{m_1}$           &  Uniform $[0.1,1]$  \\
				$S^z_1, S^z_2$  & Spin parameters of two black holes      &  Uniform $[-0.99, 0.99]$  \\
				$\iota$, $\psi$    & The inclination angle and polarization angle &  Uniform $[0, \pi]$  \\
				$\phi_c$   & Coalescence phase.&  Uniform $[0,2\pi]$  \\
				$\lambda$  & Ecliptic longitude& Uniform $[0,2\pi]$  \\
				$\beta$    & Ecliptic latitude& Uniform $[0,\pi]$  \\
				\bottomrule
			\end{tabular}
		}
		
		\subtable[Parameter space of EMRIs dataset]{
			\label{tab:EMRIs}
			\begin{tabular}{p{2.5cm}p{10cm}p{4cm}}
				\toprule
				Parameter  & Description\centering               & Parameter distribution \\
				\midrule
				$M$      & The mass of MBH & Uniform $[10^5-10^7]M_{\odot}$  \\
				$m$      & The mass of stellar-mass compact & Fix $[10 M_{\odot}]$ \\
				$a$       & Spin parameter of MBH& Uniform $[10^{-3}, 0.8]$  \\
				$p_0$     & Semi-latus rectum          & Uniform $[10, 16]$ \\
				$e_0$     & Eccentricity               & Uniform $[10^{-3},0.4]$  \\
				$\iota_0$     &The cosine of the orbit’s inclination angle from the equatorial plane & Uniform $[-0.98, 0.98]$\\
				$\theta_S$, $\theta_K$& The polar angles describing the sky location and the orientation of the spin angular momentum vector of the MBH & Uniform $[10^{-3}, \pi]$\\
				$\theta_S$, $\phi_K$  & The azimuthal angles describing the sky location and the orientation of the spin angular momentum vector of the MBH & Uniform $[10^{-3}, 2\pi]$ \\
				$\Phi_{\varphi,0}, \Phi_{\theta,0}, \Phi_{r,0}$ & The phase of azimuthal, polar, and radial modes & Fix $[0]$ \\
				\bottomrule
			\end{tabular}
		}
		
		\subtable[Parameter space of GB dataset]{
			\label{tab:GB}
			\begin{tabular}{p{2.5cm}p{10cm}p{4cm}}
				\toprule
				Parameter  & Description\centering & Parameter distribution \\
				\midrule
				$f$         & Frequency            & log-Uniform $[-4, -2]$ Hz  \\
				$\dot{f}$   & The derivative of $f$   & Fix $[10^{-14}]$ \\
				$A$         & Amplitude            & Uniform $[10^{-23},10^{-21}]$\\
				$\iota_0$, $\psi$, $\phi_0$   & The inclination angle, polarization angle and initial phase & Uniform $[0, \pi]$\\
				$\lambda$& Ecliptic longitude& Uniform $[0, 2\pi]$ \\
				$\beta$& Ecliptic latitude& Uniform $[0, \pi]$  \\
				\bottomrule
			\end{tabular}
		}
	\end{table*}

	\begin{table}
	    \centering
        \caption{The waveform information of different sources.}
	    \begin{tabular}{ccc}
        \toprule
             & 20:1 extrapolation & 1:1 extrapolation\\
        \midrule
        Input tokens & 1000 & 50  \\
        Prediction tokens & 50 & 50  \\
	MBHB & 4 points/token    & 4 points/token   \\
        EMRIs& 4 points/token    & 16 points/token  \\
	GB   & 4 points/token    & 32 points/token  \\
        \bottomrule
	    \end{tabular}
	    \label{tab:seq_info}
	\end{table}

	\subsubsection{EMRIs}
	EMRIs are a kind of black hole binary system with a mass ratio of $m/M \simeq 10^{-4}-10^{-7}$ and massive black holes (MBH) that have a mass range of $M\simeq 10^5-10^7M_{\odot}$. EMRIs waveforms are able to encapsulate the properties of space-time near a massive black hole. EMRIs are among the primary detection targets for the space-based GW detectors, possessing the potential to unveil new physical phenomena \cite{babakScienceSpacebasedInterferometer2017a, maselliDetectingFundamentalFields2022, amaro-seoaneAstrophysicsLaserInterferometer2023}.
	We employ \texttt{FastEMRIsWaveforms} (\texttt{FEW}) package \cite{katzFastExtrememassratioinspiralWaveforms2021} to generate EMRIs waveforms with a sampling rate of 5s. 
    The EMRIs signals with a duration of 1 year are randomly sliced into five waveform segments containing 4,200 points for 20:1 extrapolation (or 1600 points for 1:1 extrapolation). For continuous GWs, the random slice can simulate variations in the phase and amplitude domain of the same signal, enhancing the model's generalization capability.
	The parameter space of the EMRIs dataset is shown in Table \ref{tab:EMRIs}. 
	The complexity of the EMRIs waveform is visible in Figure \ref{fig:TDI_case_EMRI}. As the eccentricity increases, there is a corresponding increase in its complexity, which becomes particularly prominent in the frequency domain.
	
	\subsubsection{Galactic binary}
	Within the Milky Way galaxy, a substantial population of binary white dwarf systems exists, posing foreground noise challenges for space-based GW detectors.
	We use the following GB model to generate GB waveforms \cite{zhangResolvingGalacticBinaries2022}:
	\begin{equation}
		\begin{aligned}
			h_+^\text{src}{ ( t )}& =\mathcal{A}(1+\cos^2\iota)\cos\Phi(t),  \\
			h_{\times}^{\text{sr}\mathrm{c}}(t)& =2\mathcal{A}\sin\iota\sin\Phi(t),  \\
			\Phi(t)& =\phi_0+2\pi f_0t+\pi\dot{f}_0t^2+\frac\pi3\ddot{f}_0t^3,  \\
			\ddot{f}_0& =\frac{11}3\frac{\dot{f}_0^2}{f_0}. 
		\end{aligned}
		\label{equ:GB}
	\end{equation}
	
	Similar to EMRIs, GB waveforms are generated with a duration of 1 year and a sampling rate of 1/15 Hz. 
	Five slices of 4,200 points for 20:1 extrapolation (or 3200 points for 1:1 extrapolation) is randomly truncated for training and inference. 
	The parameter space of the GB dataset is shown in Table \ref{tab:GB}. 
	
	\subsubsection{Detector response and TDI 2.0}
	After generating the waveform, we project it into the LISA detector \cite{katzAssessingDataanalysisImpact2022a}. For LISA, the signals will be processed with TDI combination to suppress the overpowering laser noise. The response of space-based GW detectors is more intricate compared to ground-based detectors, accounting for factors such as satellite orbits and arm-length delays. The strain induced on link 12 is:
		\begin{equation}
			\begin{aligned}
				H_{12}(t) = h^{\rm{SSB}}_{+}(t)\times \xi_{+}(\bm{\hat{u}},\bm{\hat{v}},\bm{\hat{n}}_{12})\\
                +h^{\rm{SSB}}_{\times}(t)\times \xi_{\times}(\bm{\hat{u}},\bm{\hat{v}},\bm{\hat{n}}_{12}).
			\end{aligned}
			\label{eq:response}
		\end{equation}
  
        The $\xi_{+,\times}$ refers to the antenna pattern:
        \begin{equation}
           \begin{aligned}
            &\xi_+({\bm{\hat{u}}},\bm{\hat{v}},\bm{\hat{n}}_{12})&& =\left(\bm{\hat{u}}\cdot\bm{\hat{n}}_{12}\right)^2-\left(\bm{\hat{v}}\cdot\bm{\hat{n}}_{12}\right)^2,  \\
            &\xi_{\times}(\bm{\hat{u}},\bm{\hat{v}},\bm{\hat{n}}_{12})&& =2(\bm{\hat{u}}\cdot\bm{\hat{n}}_{12})(\bm{\hat{v}}\cdot\bm{\hat{n}}_{12}),
            \end{aligned}
        \end{equation}
        where $\bm{\hat{n}}_{12}$ is the link unit vector, $\bm{\hat{u}}$ and $\bm{\hat{v}}$ represent polarization vectors defined as the opposite direction of the polar and azimuthal angles in the Solar System Barycenter (SSB) frame respectively. 
        Due to the longer arm lengths of space-based GW detectors, the influence of arm length needs to be taken into consideration. The time of transmission from spacecraft 2 is denoted as $t_2$, and after propagating over the arm length distance to reach spacecraft 1, the reception time is $t_1$,
        \begin{equation}
            t_{1}\approx t_{2} + \frac{L_{12}}{C} - \frac{1}{2c}\int_{0}^{L_{12}}
            H(x(\lambda), t(\lambda)) d\lambda,
        \label{equ:time_delay}
        \end{equation}
        where $L_{12}$ represents the arm length of the detector.
        The variable $\lambda$ describes the path of the photon. and we approximate $t_1$ to the first order as $t({\lambda})\approx t_2+\lambda/c\approx t_2+L_{12}/c.$ Due to the slow motion of the space-based GW detector, the frequency shift is given by
        \begin{equation}
            \begin{aligned}
            y_{12}(t_{1})\approx & \begin{aligned}\frac1{2\Big(1-\hat{\bm{k}}\cdot\hat{\bm{n}}_{12}(t_1)\Big)}\Big[H_{12}\left(t_1-\frac{L_{12}(t_1)}c\right.\end{aligned}  \\
            &\left.\left.-\frac{\hat{\bm{k}}\cdot\bm{x}_2(t_1)}c\right)-H_{12}\left(t_1-\frac{\hat{\bm{k}}\cdot\bm{x}_1(t_1)}c\right)\right],
            \end{aligned}
        \end{equation}
        where $\hat{k}$ represents the propagation vector of the wave source. 

	Space-based GW detectors have unequal arm lengths, which results in significant laser frequency noise. 
	To mitigate this issue, TDI techniques are commonly employed to suppress laser frequency noise \cite{tinto_time_2004, tintoTimeDelayInterferometry2014}.
    The first and second generation Michelson combinations, X1 and X2, are defined by \cite{tinto_time_2004},
    \begin{equation}
        \begin{aligned}
        X_{1}=&y_{13}+\bm{D}_{13}y_{31}+\bm{D}_{131}y_{12}+\bm{D}_{1312}y_{21}\\
        &-\left[y_{12}+\bm{D}_{12}y_{21}+\bm{D}_{121}y_{13}+\bm{D}_{1213}y_{31}\right],\\
        \end{aligned}
    \end{equation}
    \begin{equation}
        \begin{aligned}
        X_{2}=&X_1+\bm{D}_{13121}y_{12}+\bm{D}_{131212}y_{21}+\bm{D}_{1312121}y_1\\
        &+\bm{D}_{13121213}y_{31}-[\bm{D}_{12131}y_{13}+\bm{D}_{121313}y_{31}\\
        &\left.+\bm{D}_{1213131}y_{12}+\bm{D}_{12131312}y_{21}\right],
        \end{aligned}
    \label{equ:tdi2.0}
    \end{equation}
    where the delay operators are defined by,
    \begin{equation}\bm{D}_{i_1,i_2,...,i_n}x(t)=x\Bigg(t-\sum_{k=1}^{n-1}L_{i_ki_{k+1}}(t)\Bigg).\end{equation}
	The detector response and TDI 2.0 response of GW are calculated using \texttt{Fastlisaresponse} \cite{katzAssessingDataanalysisImpact2022a}. 
	TDI 2.0 generates three channels X, Y, and Z. The variables Y and Z may be produced via cyclic permutation of the indices in Eq. \ref{equ:tdi2.0}. A more detailed derivation can be found in Section IV of the reference \cite{katzAssessingDataanalysisImpact2022a}.
	By combining X, Y, and Z, three independent channels A, E, and T are obtained, 
	\begin{equation}
		\begin{aligned}
			A &= (Z-X)/\sqrt{2},\\
			E &= (X-2Y+Z)/\sqrt{6},\\
			T &= (X+Y+Z)/\sqrt{3}.
		\end{aligned}
	\end{equation}
	
    The incorporation of response functions and TDI 2.0 combination introduces increased complexity to the waveform, especially in the high-frequency part.
	As depicted in Figure \ref{fig:TDI_complex}, MBHB waveforms exhibit significant differences at various parameter values.
	
	\subsection{CBS-GPT Model}
	Transformers \cite{vaswaniAttentionAllYou2017} are a class of deep learning models that have exhibited excellent performance in various tasks, such as NLP \cite{Brown2020LanguageMA} and CV \cite{Liu2021}. 
	We incorporate the masked self-attention mechanism and feed-forward neural network to build our CBS-GPT model.
	
	\textbf{Patching.}
	Firstly, the input waveform is preprocessed by standardization, which facilitates the model in capturing waveform information more effectively: 
	\begin{equation}
		I = \text{standard} (s) = \frac{s - \text{mean}(s)}{\text{std}(s)}
	\end{equation}
 where $s=\{s_i|i\in [0, N)\}$ represents the input waveform, $\mu=\text{mean}(s)=\frac{1}{N}\sum_{i=1}^N s_i$ and $\text{std}(s)=\sqrt{\frac{\sum_{i=1}^N (s_i -\mu)^2}{N}}$ represent the mean and standard deviation of the waveform, respectively. 
        The standardization centers the original data to a mean of 0 and a standard deviation of 1, which makes features have equal weight in various analyses and is more suitable for machine learning algorithms that are sensitive to feature scales.
        Then, $I=\{x_i|i\in [0, N)\}$ is divided into non-overlapping patches, and we refer to each patch as a "token" here. 
        In our 20:1 extrapolation experiment for example, we have an input waveform with $N=4200$ sampling points, which is segmented into $num=1,050$ tokens, and each token contains $4$ points.
    Each token is treated as a vector, after patching, the standardized waveform $I$ is processed into the input matrix $I^{'} \in \Bbb R^{[1050, 4]}$.
	
	\textbf{Hybrid Embedding.}
	The hybrid embedding module is utilized in our model, because each token contains richer physical information and cannot be tokenized by simple tokenizers as in NLP.
	As Figure \ref{fig:GPT-model} shows, it is combined with a token embedding layer and a positional embedding layer (Eq. \ref{eq:hybrid}).
    The token embedding layer performs linear projection to achieve dimension-matching with following encoder blocks, which meanwhile preserves the entire information of the input waveform. 
    The positional embedding is also a linear layer that encodes positional relationships between tokens, which is rather important in improving prediction accuracy \cite{vaswaniAttentionAllYou2017}. 
           
    \begin{equation}
		\begin{aligned}
			&E_e = I{'}\ W_e\\
			&E_p = \mathbf I_{num}\ W_p\\ 
			&E_{hybrid} = E_e + E_p,
		\end{aligned}
        \label{eq:hybrid}
	\end{equation}
    where $W_e\in\Bbb R^{[4, d_{model}]}$, $d_{model}=2048$ and $W_p \in \Bbb R^{[num, d_{model}]}$ are both learnable parameters, and $\mathbf I_{num}$ represents an identity matrix with shape $[num, num]$. 
    
	\textbf{Encoder block.}
	The encoder contains $n_{block}=36$ blocks. Each block mainly consists of an attention module and a feed-forward neural network.
	As for the attention module, masked multi-heads self-attention (MMHSA) is adopted in our work, which enables information to be projected into matrices in different ways, thereby enhancing the expressive capacity of the model. 
	The computation process of the attention module is as follows: 
		\begin{equation}
			\begin{aligned}
                &Q_{ji}=W_{ji}^Q x_j \\
                &K_{ji}=W_{ji}^{K}x_{j} \\
                &V_{ji}=W_{ji}^{V}x_{j} 
                \end{aligned}
		\end{equation}
            \begin{equation}
				head_{ji}(Q_{ji},K_{ji},V_{ji})=\text{softmax}\left(\frac{Q_{ji}K_{ji}^T \cdot \mathrm{mask}}{\sqrt{d}}\right)V_{ji}, 
		\end{equation}
  	\begin{equation}
		\mathrm{MMHSA}_j(Q,K,V)=\mathrm{Concat}(head_1,...,head_H)W_j^E \:, 
	\end{equation}
	\begin{equation}
		\mathrm{H_j^{\prime}}=\mathrm{LayerNorm}(\mathrm{MMHSA}(Q,K,V))+x_j \:,
	\end{equation}
 
        In each encoder block, there is $H=d_{model}/64=32$ heads. $W_{ji}^Q, W_{ji}^{K}, W_{ji}^{V}$ represent learnable query, key, and value parameters of $i$-th attention head and $j$-th encoder block, respectively, and $mask$ is a lower triangular standard matrix.
 	\begin{equation}
            x_j=
		\begin{cases}
                E_{hybrid}, & j=0 \\
                y{'}_{j-1}, & 0<j<n_{block},
            \end{cases}
	\end{equation}
	where $x_j$ represents the hybrid embedding or the output of the previous encoder block. 
        The feed-forward network (\textbf{FFN}) is composed of two dense layers  and is connected to each attention module.
	
 We employ the residual connection (Eq. \ref{eq:resd}), which is helpful to alleviate the gradient-vanishing problem.
	\begin{equation}
		\mathrm{Inter}(H_j^{\prime})=\mathrm{GeLU}(H_{j}^{\prime}W_{j1}+b_{j1})W_{j2}+b_{j2},
	\end{equation}
	\begin{equation}
		y_{j}=\mathrm{FFN}(H_j^{\prime})=\mathrm{LayerNorm}(\mathrm{Inter}(H_j^{\prime}))+H_j^{\prime},
        \label{eq:resd}
	\end{equation}
        where $W_{j1},b_{j1}, W_{j2}, b_{j2}$ are both learnable parameters and $GeLU$ is an activation function. 
        Finally, the output of the last encoder block is inversely projected to the same shape of $I{'}$.
        \begin{equation}
            y' = y_{n_{block}} W_i^T,\ y' \in \Bbb R^{[1050, 4]}.
        \end{equation}
	
	\textbf{Loss Function.}
        Next-token-prediction error is adopted to train CBS-GPT, which means that the predicted token $y{'}_{m}$ is designed to match the input token ($I{'}_{m+1}$) at position $m+1$.
    Hence, only $num-1$ tokens are taken into account when calculating the training loss.
        Specifically, the mean squared error (MSE) loss is used to measure the difference between the predictions:
	\begin{equation}
		\mathcal{L}=\frac {1} {(num-1) \times 4} \sum_{m=0}^{num-2} \sum_{t=0}^3 || y{'}_{m,t}-I{'}_{m+1,t}||^2
		\label{equ:loss_function}
	\end{equation}
	
\subsection{Training and Inference}\label{subsec:tni}

	During training, the Adam \cite{Kingma2017} optimizer with $\beta_1=0.9$, $\beta_2=0.999$ is used, and the initial learning rate is 2e-4.
    There are 1.6 millions waveforms in the training dataset of each model, and the parameter of each waveform is randomly selected from its correponding parameter space.
	After passing through the LISA response, each waveform is divided into three TDI channels (A, E, and T). 
	In this study, the E channel is selected to train the model.
	The model was trained on two NVIDIA V100 GPUs for approximately 30 hours.
 	During inference, for each signal source, 10,000 waveforms are generated to test CBS-GPT’s performance.
For each waveform, the initial input contains 1,000/50 valid tokens and 50 masked tokens that are masked with zero, whose corresponding value in the mask matrix also equals zero, which guarantees that no attention is paid to the to-be-extrapolated token.
	In the first step, the 1,001-st/51-st token is predicted and replaces the previous 1,001-st/51-st token, and so forth, 50 successive tokens are predicted based on 1,000/50 valid input tokens.
	
\section{Results and discussion}
\label{sec:res}
	During inference, overlap is defined to evaluate the extrapolation accuracy of the predicted waveform. 
	Overlap is calculated between the target waveform and the predicted waveform generated by CBS-GPT as stated in Eq. \ref{equ:overlap}.
	The overlap $\mathcal{O}$ ranges between $[0,1]$, with values closer to 1 indicating that the predicted waveform is more similar to the target waveform.
	
	\begin{equation}
		\mathcal{O}({h}_t, {h}_p)=
		\max_{t_c}\left(\hat{h}_t|\hat{h}_p[t_c]\right)^{1/2} ,
		\label{equ:overlap}
	\end{equation}
	with
	\begin{equation}
		\begin{aligned}
			(h|s) &=2\int_{f_{\min}}^{f_{\max}}\frac{\tilde{h}^*(f)\tilde{s}(f)+\tilde{h}(f)\tilde{s}^*(f)}{S_n(f)}df, \\
			\hat{h} &= \frac{h}{\sqrt{(h|h)}}
		\end{aligned}
	\end{equation}
	where $t_c$ represents time-shifted, and we set $S_n(f)=1$.
 
	Overall, in the context of 20:1 extrapolation tasks targeting MBHB, GB, and EMRIs signals, CBS-GPT has demonstrated remarkable efficacy, with over 50\% of the overlaps exceeding 0.99. 
     Figure \ref{fig:mbhb_overlap} and \ref{fig:CGW_overlap} showcase the prediction performance of each waveform under varying parameter conditions, revealing that the CBS-GPT model can learn waveform features with a wide range of parameters.
     Figure \ref{fig:showcase} and \ref{fig:interprete} demonstrate the generalization and potent interpretability  of CBS-GPT.

\subsection{Results of MBHB}
\label{subsubsec:mbhb_result}

    The results of MBHB overlap are shown in Table \ref{tab:overlap_MBHB}.
    The CBS-GPT model is sensitive to total mass, mass ratio, and spin parameters. Here we use $\chi_{\text{eff}} $ to represent the spin parameter \cite{barack_black_2019}: 
	\begin{equation}
		\chi_{\text{eff}} = \frac{S_{1}^z}{1+q} + \frac{qS_{2}^z}{1+q}.
		\label{eq:spin_eff}
	\end{equation}
    
\begin{figure*}
        \centering
        \subfigure[MBHB overlap distribution.]{
            \label{fig:mbhb_overlap:hist}
            \includegraphics[width=0.21\textwidth]{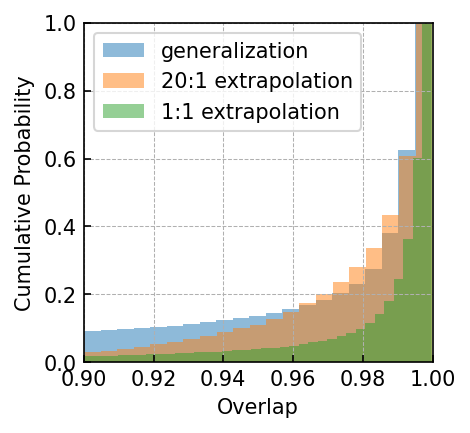}
            }
        \subfigure[MBHB: 20:1  extrapolation.]{
            \label{fig:mbhb_overlap:1000by50}
            \includegraphics[width=0.22\textwidth]{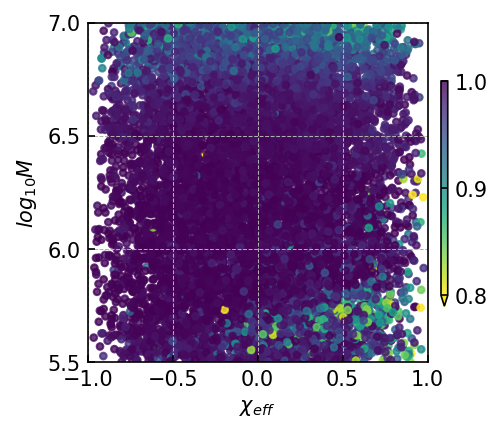}
            }
	\subfigure[MBHB: 1:1  extrapolation]{
		  \label{fig:MBHB_overlap_50by50}
            \includegraphics[width=0.22\textwidth]{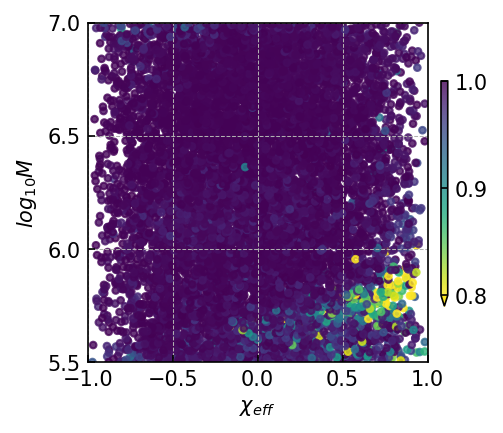}
            }
        \subfigure[MBHB generalization: 1:1 extrapolation]{
            \label{fig:MBHB_generalization}
            \includegraphics[width=0.22\textwidth]{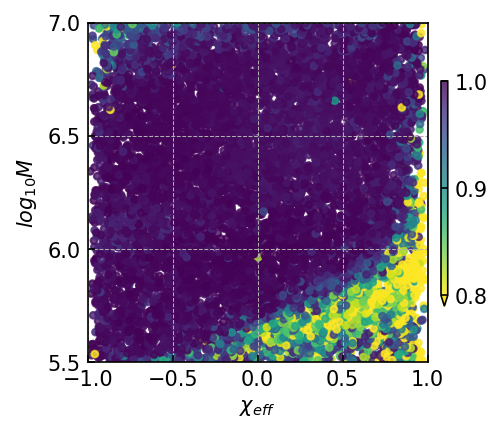}
            }
        \caption{\textbf{The overlap distribution of MBHB }is shown in \textbf{(a)}. 
        \textbf{(b, c, d)} portray the heat maps of $M_{tot}$ and $\chi_{\mathrm{eff}}$ parameters, which have the greatest impact on overlap. A darker color corresponds a higher overlap value.}
	\label{fig:mbhb_overlap}
	\end{figure*}

    \textbf{20:1 extrapolation.}
	The overlap distribution and waveform examples are shown in Figure \ref{fig:mbhb_overlap:hist} and Figure \ref{fig:showcase_mbhb_20:1}, with mean and median overlaps equal 0.981 and 0.992, respectively.
	The overlap results reveal that CBS-GPT can forecast the waveform of the merge-ringdown phase based on the inspiral phase characteristics. 
	CBS-GPT exhibits optimal inference performance when the total mass is approximately $10^{6.5}M_{\odot}$ as shown in Figure \ref{fig:mbhb_overlap:1000by50}. 
	This phenomenon has also been observed in other signal sources. 
 	The overlap is lower for for waveforms with low total mass and high effective spin $\chi_{\text{eff}} $. 
        Comparing low and high-mass situations to those involving intermediate masses, the performance of mid-frequency band prediction is the best.
	Since TDI 2.0 transfer functions in the high-frequency part are more complex \cite{larson_sensitivity_2000, babak_lisa_2021}, the waveform is also more complex. Consequently, the model's performance experiences a slight decrease.    
    But even under such less ideal circumstances, CBS-GPT can still successfully recover a significant portion of the signals. 

    \textbf{1:1 extrapolation.} 
    We find that the model pays little attention to the early-stage waveform and mainly concentrates on the late-stage inspiral waveform when forecasting the merging waveform of an MBHB (detailed explanation is in Section \ref{sec:res_attention_map}). This demonstrates the marginal contribution of early-stage inspiral waveforms to subsequent waveforms generation.
    Hence we retrained a model, whose input only contains 200 points before merge time and predicted the subsequent 200 points, thus achieving a 1:1 extrapolation. The average and median overlap achieved 0.990 and 0.996, respectively. The results are slightly better than the previous 20:1 extrapolation, which validates our former conclusion. 
    In Table \ref{tab:overlap_MBHB}, we observe a noticeable improvement in overlap for cases with masses greater than $10^6 M_\odot$, which illustrates that shorter input waveforms allow the model's attention to be more focused, leading to improved inference performance. 
    In Figure \ref{fig:showcase_mbhb_1:1}, we showcase the predictive performance of CBS-GPT in the 1:1 extrapolation scenario.
    
    \textbf{Generalization ability} refers to the performance of a model when applied to data that has not seen before. 
    To evaluate the generalization capability of CBS-GPT, we selected MBHB signals with mass ratios ranging from 1:10 to 1:100  in the 1:1 extrapolation model. Figure \ref{fig:showcase_gen} showcases the waveform examples of generalization ability.
    The average overlap achieved 0.970, with more than half of the overlaps surpassing 0.993, which demonstrated the strong generalization ability of our method. 
    The model's performance on generalization experiment also illustrates its ability to learn the essence of the data.

    \begin{table}
    \caption{\textbf{The overlap results}}
    \subtable[\textbf{The overlap results of MBHB.} 
    Group iii. and iv. correspond to the mean overlap values for $M_{tot}<10^6 M_\odot$ and $M_{tot}\geq 10^6 M_\odot$ respectively. ]{
    \label{tab:overlap_MBHB}
    \begin{tabular}{p{3cm}p{1cm}<{\centering}p{1cm}<{\centering}p{2cm}<{\centering}}
        \toprule
        MBHB & 20:1 & 1:1 & generalization  \\
        \midrule
         i. mean        & 0.981 & 0.990 & 0.970  \\
         ii. median     & 0.992 & 0.996 & 0.993  \\
         iii. mass $< 10^6 M_\odot$ & 0.980 & 0.979 & 0.938  \\
         iv. mass $\geq 10^6 M_\odot$ & 0.982 & 0.995 & 0.986  \\
         \bottomrule
    \end{tabular}}
        
    \subtable[\textbf{The overlap results of EMRIs.} Group iii. and iv. correspond to the mean overlap values for $e_0<0.1$ and $e_0\geq 0.1$ respectively.]{
    \label{tab:EMRI}
    \begin{tabular}{p{3cm}p{2cm}<{\centering}p{2cm}<{\centering}}
        \toprule
        EMRIs & 20:1 & 1:1 \\
        \midrule
         i.mean        & 0.912 & 0.807  \\
         ii. median    & 0.997 & 0.910  \\
         iii. $e_0 < 0.1$  & 0.962 & 0.905  \\
        iv. $e_0 \geq 0.1$ & 0.896 & 0.778  \\
         \bottomrule
        \label{tab:emri_overlap}
        \end{tabular}}

    \subtable[\textbf{The overlap results of GB.} Group iii. and iv. correspond to the mean overlap values for $f < 10^{-3}$ and $f \geq 10^{-3}$ respectively.]{
    \label{tab:GB-2}
    \begin{tabular}{p{3cm}p{2cm}<{\centering}p{2cm}<{\centering}}
        \toprule
        GB & 20:1& 1:1  \\
        \midrule
         i. mean       & 0.991 & 0.992  \\
         ii. median    & 0.996 & 0.994  \\
         iii. $f < 10^{-3}$   & 0.987 & 0.990  \\
         iv. $f \geq 10^{-3}$ & 0.995 & 0.993  \\
         \bottomrule
        \end{tabular}}
    \label{tab:continous_result}
    \end{table}

    \subsection{Results of Continous Waveform: EMRIs and GB}
\begin{figure*}
        \centering
        \subfigure[EMRIs overlap distribution.]{
            \label{fig:emri_overlap:hist}
            \includegraphics[width=0.22\textwidth]{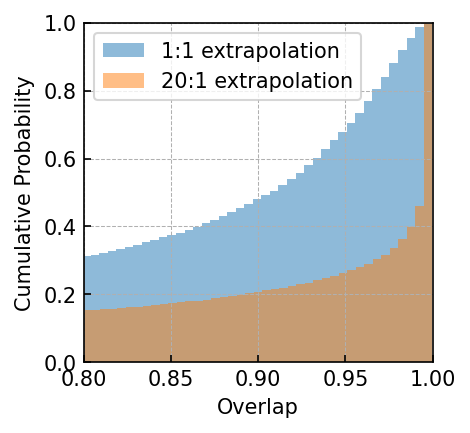}
            }
        \subfigure[EMRIs: 20:1  extrapolation.]{
            \label{fig:emri_overlap:1000by50}
            \includegraphics[width=0.22\textwidth]{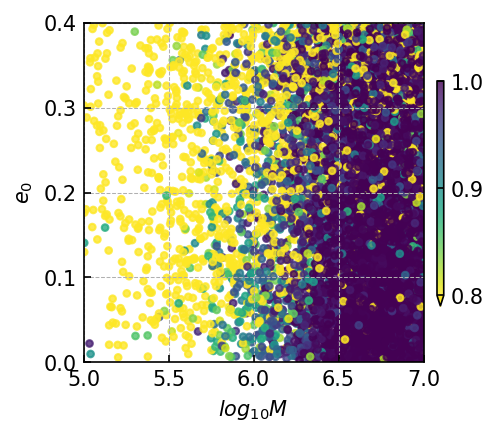}
            }
	\subfigure[EMRIs: 1:1  extrapolation]{
             \label{fig:emri_overlap_50by50}
             \includegraphics[width=0.22\textwidth]{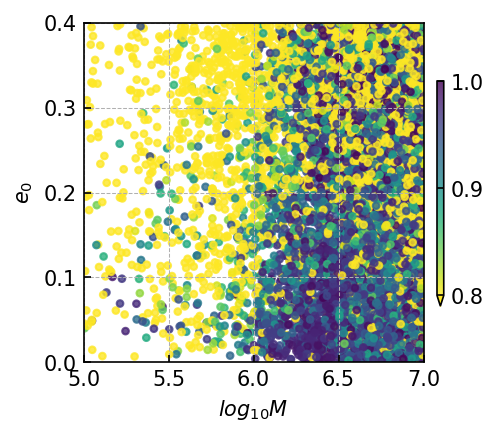}
            }

        \subfigure[GB overlap distribution.]{
            \label{fig:gb_overlap:hist}
            \includegraphics[width=0.22\textwidth]{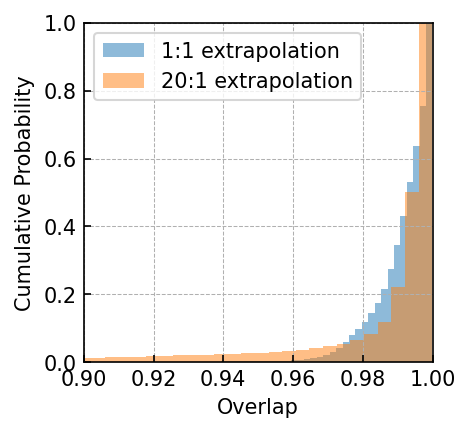}
            }
        \subfigure[GB: 20:1  extrapolation.]{
            \label{fig:gb_overlap:1000by50}
            \includegraphics[width=0.22\textwidth]{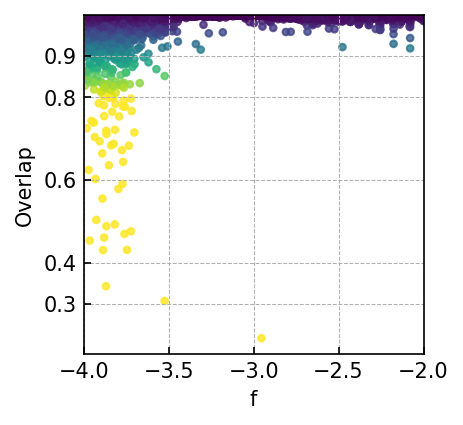}
            }
	\subfigure[GB: 1:1  extrapolation]{
             \label{fig:gb_overlap_50by50}
             \includegraphics[width=0.22\textwidth]{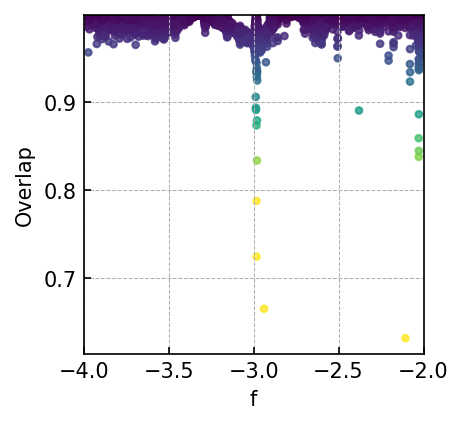}
            }
        
        \caption{\textbf{The overlap distributions of EMRIs and GB }are shown in \textbf{(a, d)}.  
        \textbf{(b, c)} portray the heat maps of $e_0$ and $M$ parameters, which have the greatest impact on overlap of EMRIs. 
        Similarly, \textbf{(e, f)} portray the heat maps of frequency parameter $f$, which have the greatest impact on overlap of GB. 
        A darker color corresponds a higher overlap value.}
	\label{fig:CGW_overlap}
    \end{figure*}

\begin{figure*}
        \centering
        \subfigure[MBHB 20:1 extrapolation]{\label{fig:showcase_mbhb_20:1}
        \includegraphics[width=0.32\textwidth]{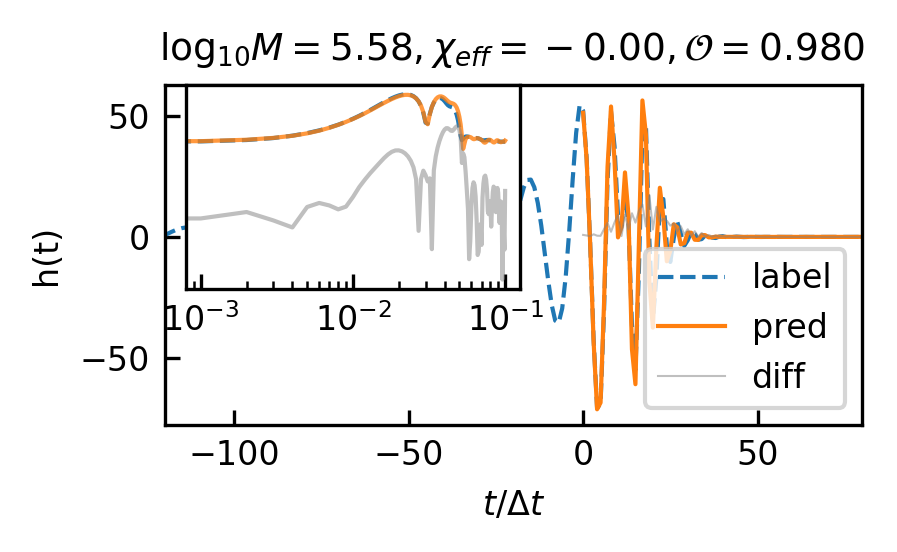}}
        \subfigure[MBHB 1:1 extrapolation]{\label{fig:showcase_mbhb_1:1}
        \includegraphics[width=0.32\textwidth]{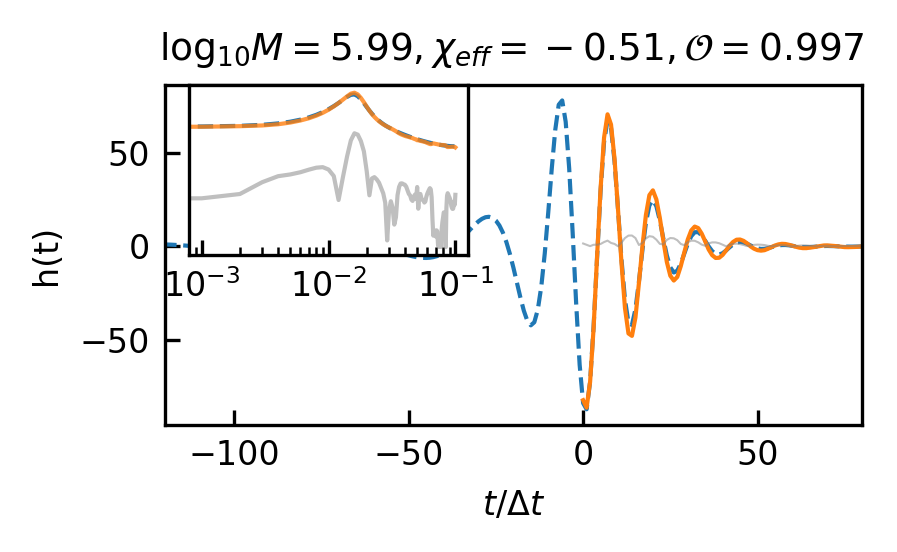}}
        \subfigure[EMRIs 20:1 extrapolation]{\label{fig:showcase_emri_20:1}
        \includegraphics[width=0.32\textwidth]{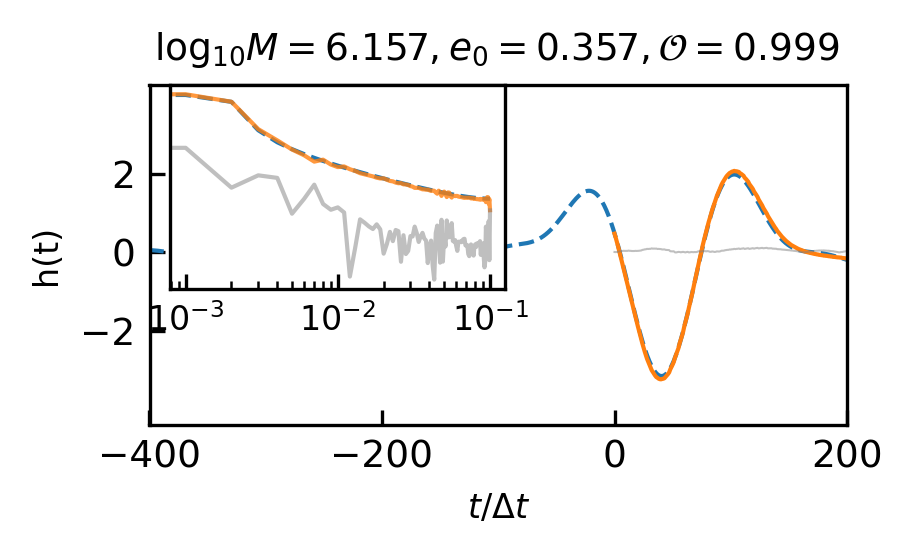}}
        \subfigure[EMRIs 1:1 extrapolation]{\label{fig:showcase_emri_1:1}
	\includegraphics[width=0.32\textwidth]{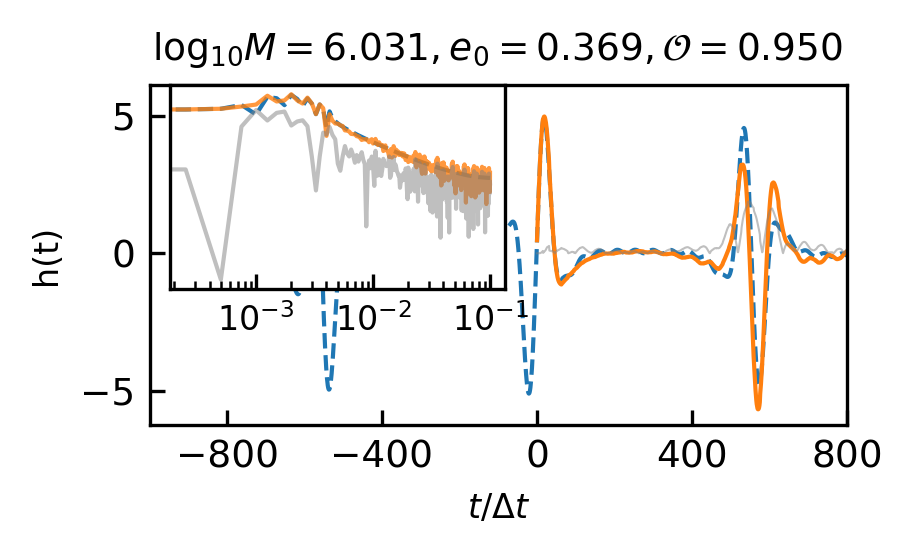}}
        \subfigure[GB 20:1 extrapolation ]{\label{fig:showcase_gb_20:1}
        \includegraphics[width=0.32\textwidth]{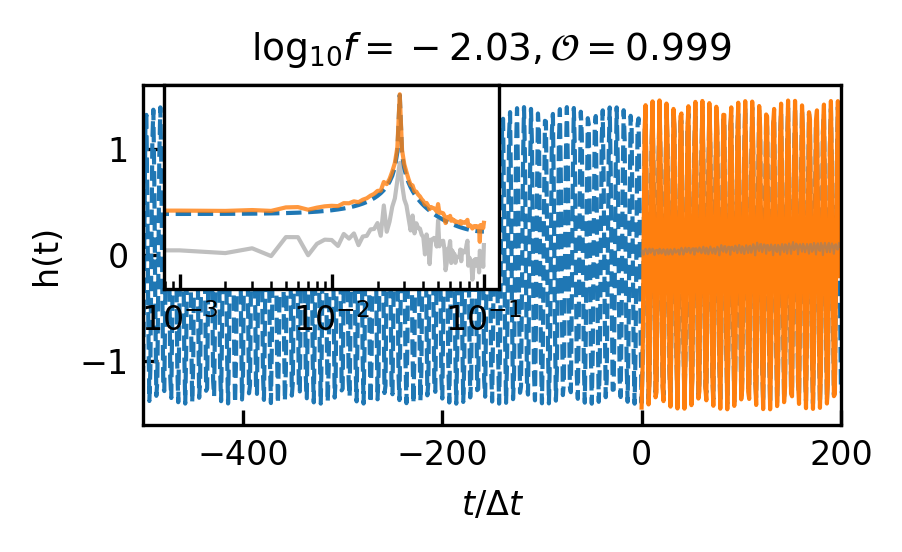}}
        \subfigure[GB 1:1 extrapolation]{\label{fig:showcase_gb_1:1}
        \includegraphics[width=0.32\textwidth]{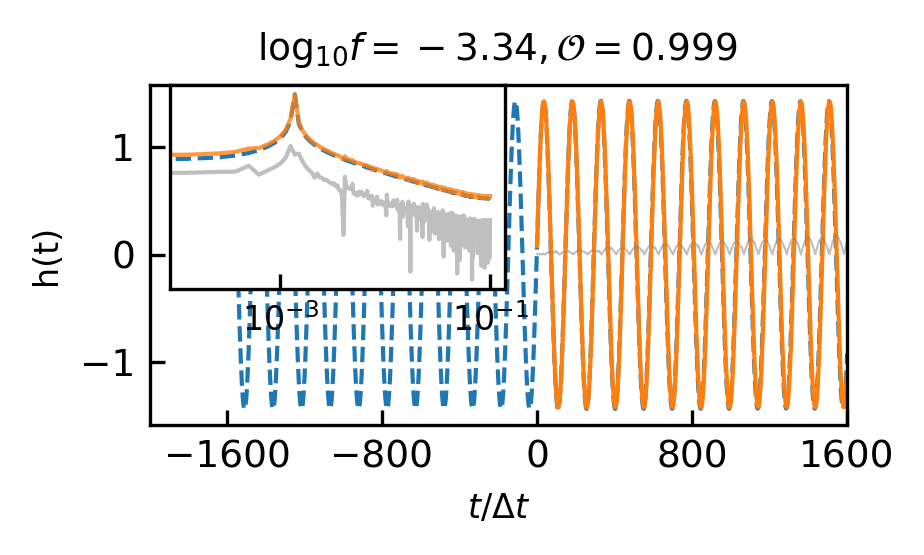}}
        \subfigure[\textbf{MBHB Generalization 1:1 extrapolation.}]{\label{fig:showcase_gen}
        \includegraphics[width=0.7\textwidth]{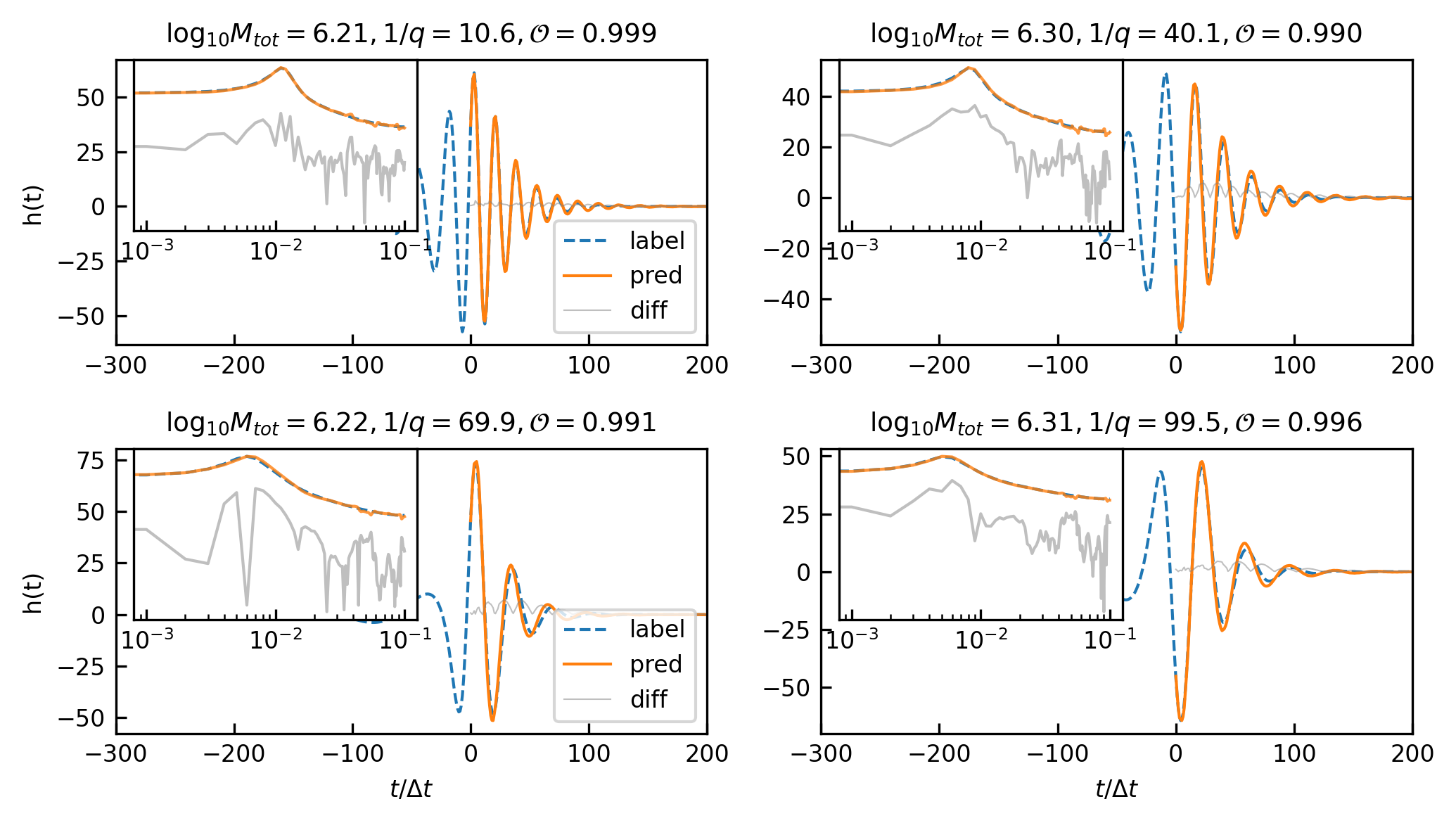}}
        \caption{\textbf{CBS-GPT prediction results. (a, b) MBHB results. (c, d) EMRIs results. (e, f) GB results. (g) Generalization results of MBHB waveform with $1/q \approx$ 10, 40, 70, and 100, respectively.} We set the predicted starting point at time zero. The blue line represents the conjunction of the last part of the input waveform and target label, the orange line is the predicted waveform, and the gray line is the difference between the predicted and target waveform. The inset figure in each subfigures represents the anticipated and target waveforms in the frequency domain, as well as the differences between them. } 
        \label{fig:showcase}
    \end{figure*}

    The overlap distributions of the EMRIs and GB are shown in Figure \ref{fig:CGW_overlap} and their mean and median values are displayed in Table \ref{tab:EMRI} and Table \ref{tab:GB-2}. 
    Examples of predicted EMRIs and GB waveforms are shown in Figure \ref{fig:showcase_emri_20:1}-\ref{fig:showcase_gb_1:1}.
 
    \textbf{20:1 extrapolation.} Regarding GB, its mean and median overlap both exceed 0.99.
	The mean and median overlap of EMRIs are equal to 0.912 and 0.997, respectively. 
	While the mean overlap of EMRIs is slightly lower, its median overlap aligns with that observed in MBHB and GB waveforms. 
	
	Specifically, the overlap distribution of EMRIs significantly influenced by the mass parameters and eccentricity parameters. 
	As depicted in Table \ref{tab:emri_overlap}, when $e_0$ is less than $0.1$, the majority of overlaps remain below 0.9. 
	As the eccentricity increases, the waveform features become more complex in waveform amplitude. 
	Therefore, when the eccentricity is higher, the corresponding overlap tends to decrease. 
	
	In contrast to MBHB and EMRIs signals, the GB signal presents a comparatively straightforward, single-frequency waveform. 
	As for GB, the frequency parameter has the greatest impact on the waveform. 
	When the frequency is larger than $10^{-3.5} $Hz, the overlap is basically higher than 0.9. 
	The result of GB signals demonstrates the model's sensitivity over frequency, with the distinct preference for learning the characteristics associated with intermediate frequency signals.

    \textbf{1:1 extrapolation.} 
    In this scenario, the mean and median overlaps for EMRIs were found to be 0.807 and 0.910, while for GB, the mean and median overlaps were 0.992 and 0.994 respectively.
    
    The performance impact was negligible for GB, but there was a significant decrease in EMRIs waveforms. This can be attributed to the larger eccentricity and wider range of scales exhibited by EMRIs, as well as their continuous periodic transitions. 
    Due to the high complexity of EMRI waveforms, shorter waveforms fail to capture the waveform features. Therefore, in the case of complex waveforms, CBS-GPT requires longer input waveforms to learn more distinctive features. 

	\subsection{Interpretability}
	\label{sec:res_attention_map}
\begin{figure*}
        \centering
        \subfigure[MBHB 20:1 extrapolation, $M_{tot}=10^{5.5} M_\odot$]{\label{fig:interprete_a}
        \includegraphics[width=0.3\linewidth]{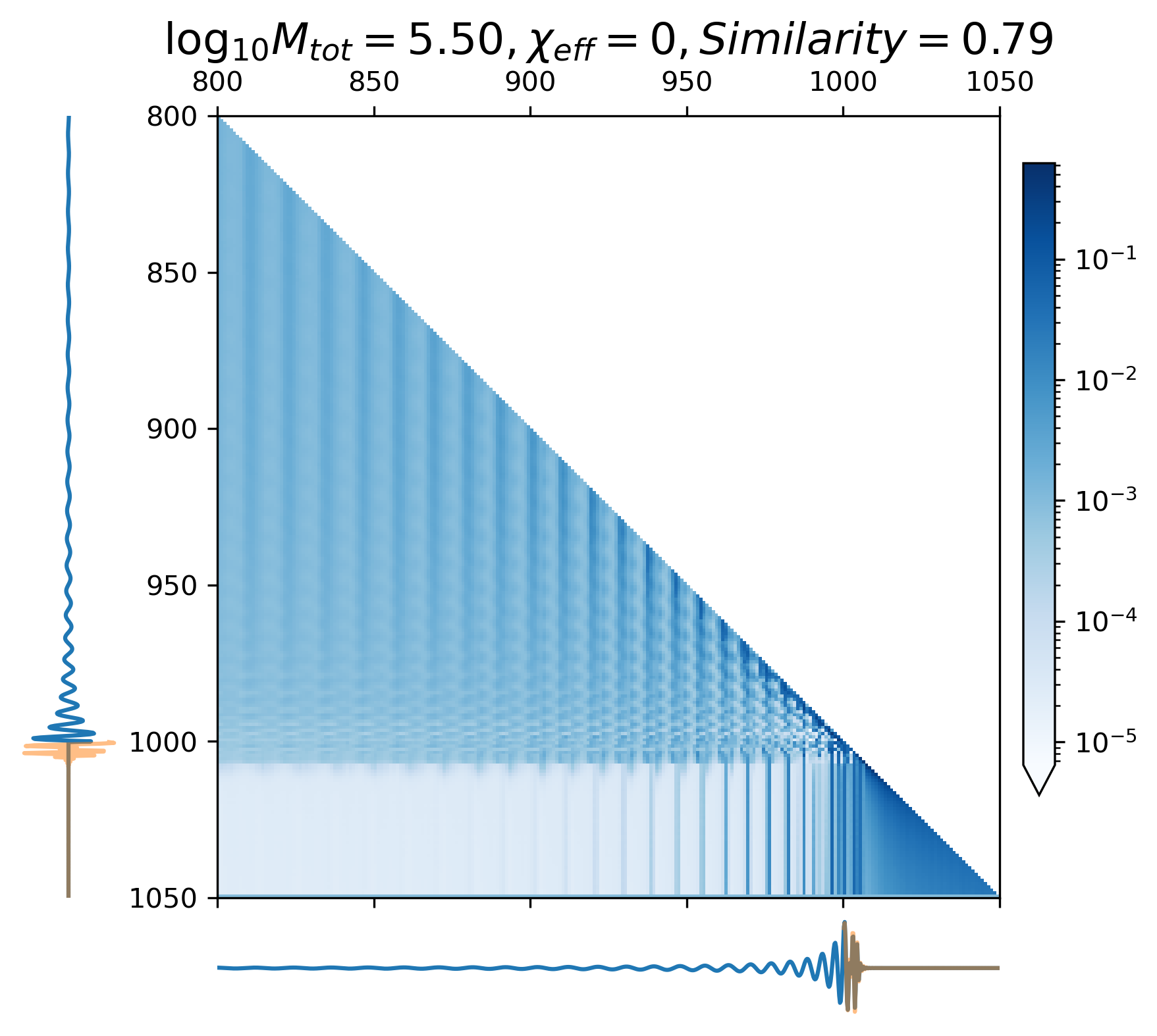}}
        \subfigure[MBHB 20:1 extrapolation, $M_{tot}=10^{6.25} M_\odot$]{\label{fig:interprete_b}
        \includegraphics[width=0.3\linewidth]{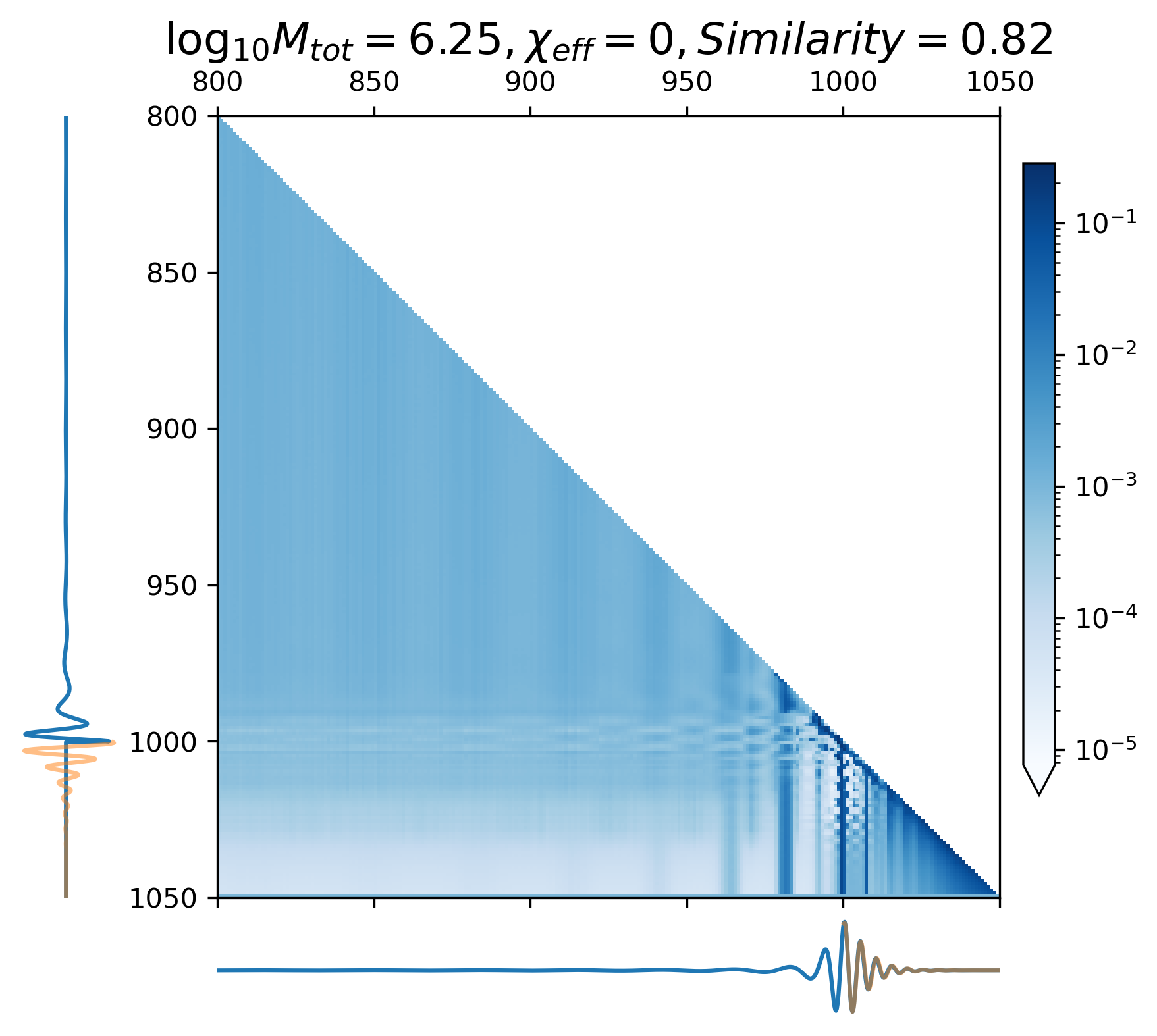}}
        \subfigure[MBHB 1:1 extrapolation, $M_{tot}=10^{6.25} M_\odot$]{\label{fig:interprete_c}
        \includegraphics[width=0.3\linewidth]{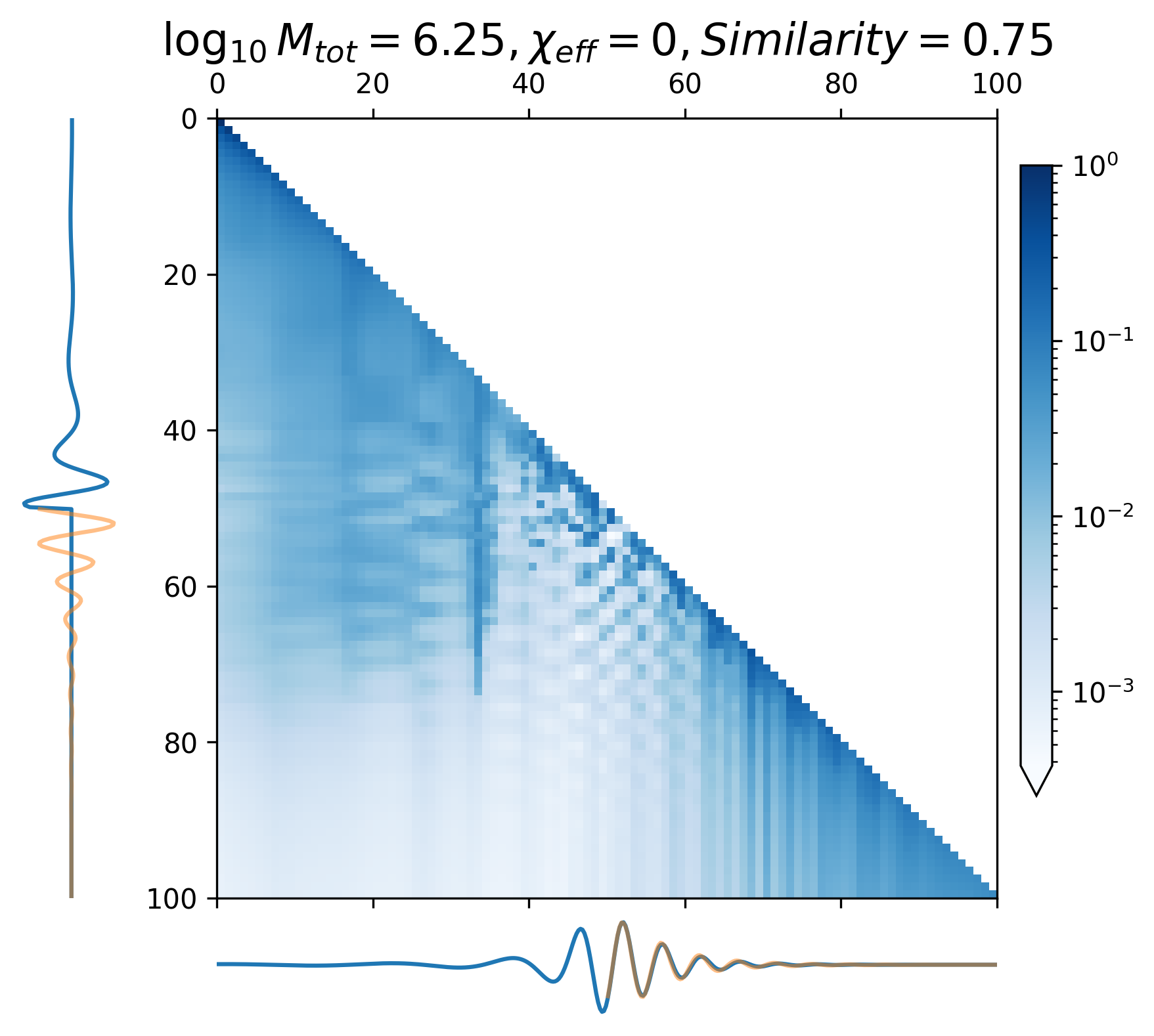}}
        \subfigure[EMRIs 20:1 extrapolation: $e_0=0.01$]{\label{fig:interprete_d}
	\includegraphics[width=0.3\linewidth]{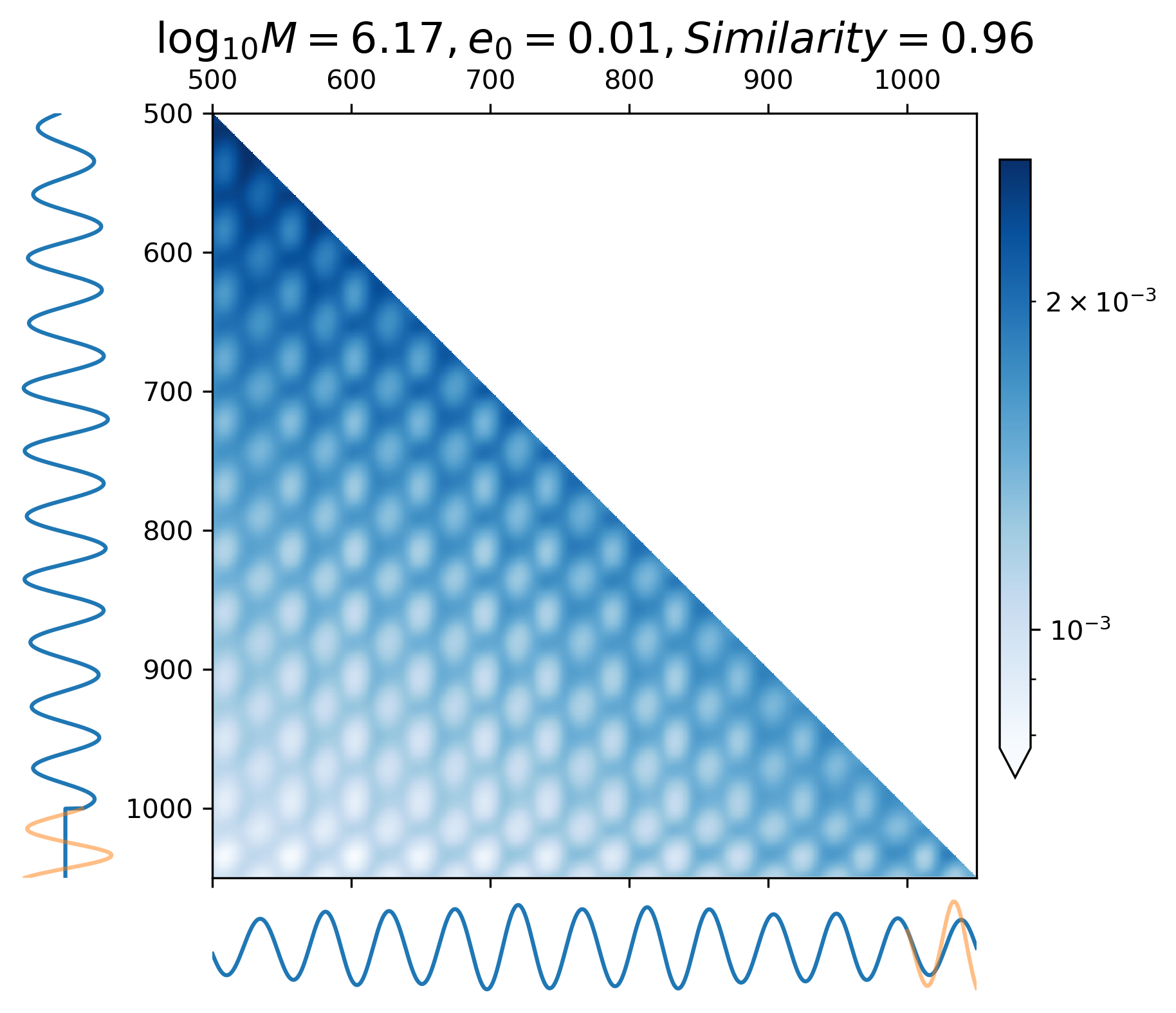}}
        \subfigure[EMRIs 20:1 extrapolation: $e_0=0.3$ ]{\label{fig:interprete_e}
        \includegraphics[width=0.3\linewidth]{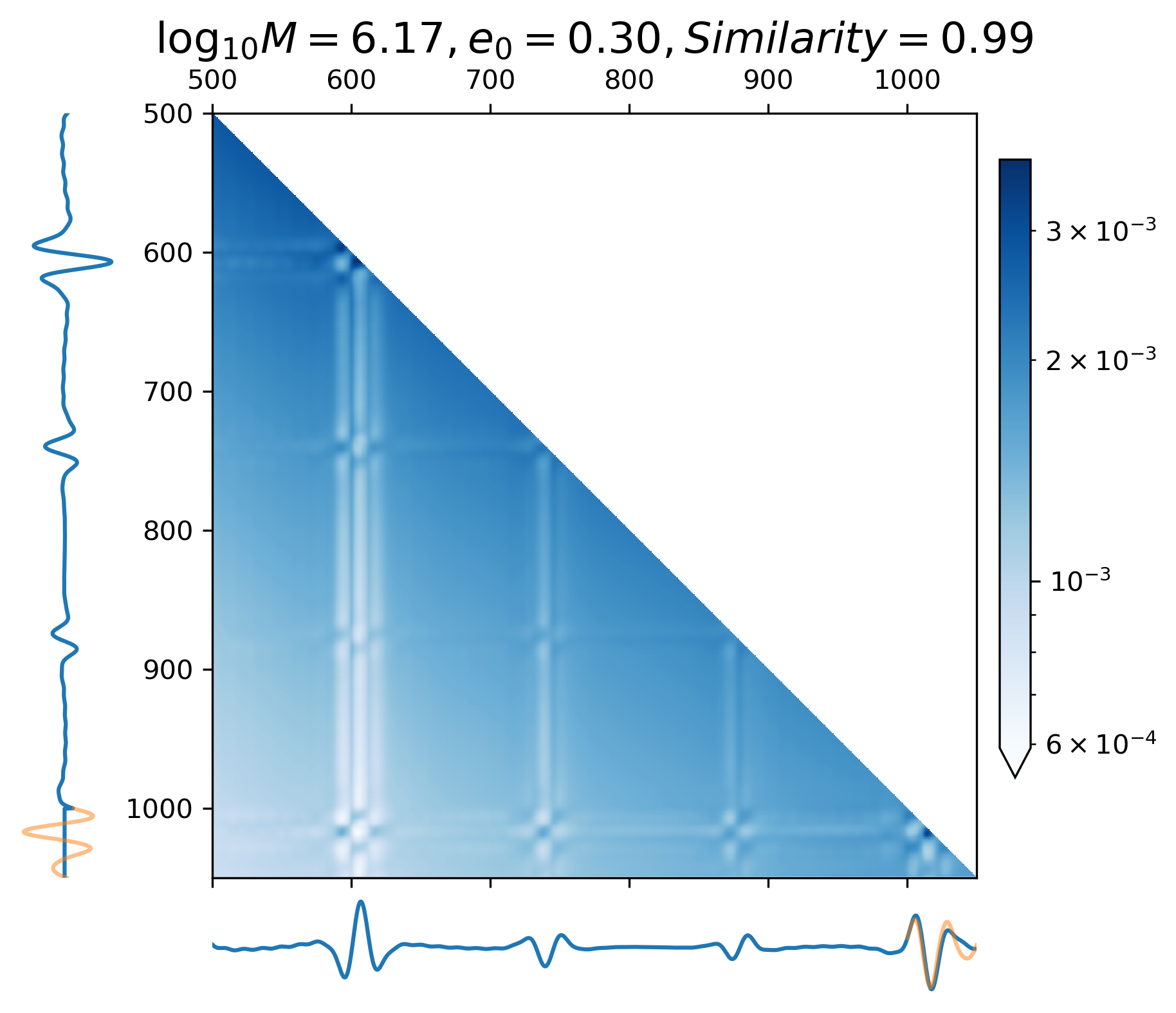}}
        \subfigure[EMRIs 1:1 extrapolation: $e_0=0.3$ ]{\label{fig:interprete_f}
        \includegraphics[width=0.3\linewidth]{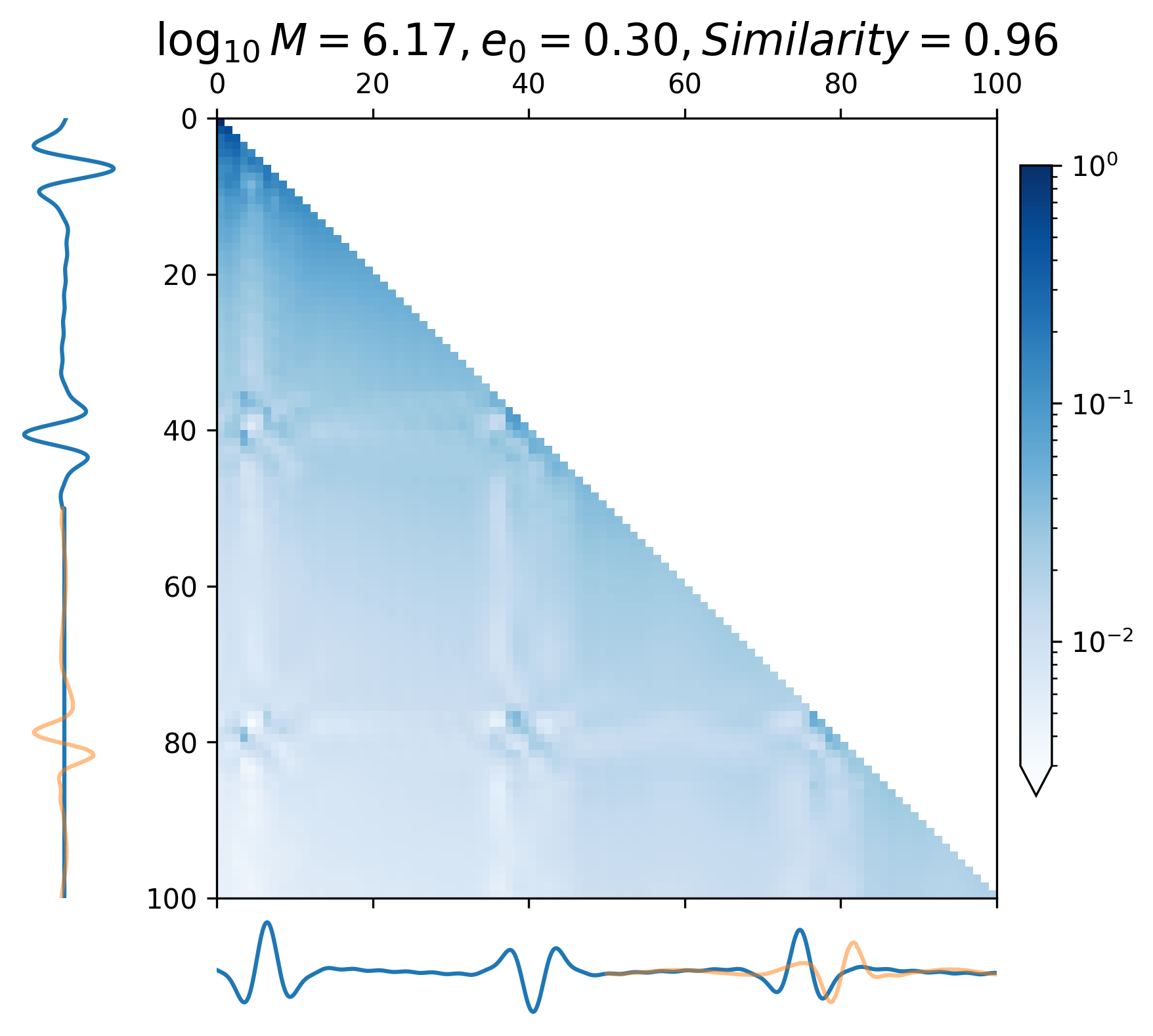}}
        \subfigure[GB 20:1 extrapolation: $f=10^{-3}$ ]{\label{fig:interprete_g}
        \includegraphics[width=0.3\linewidth]{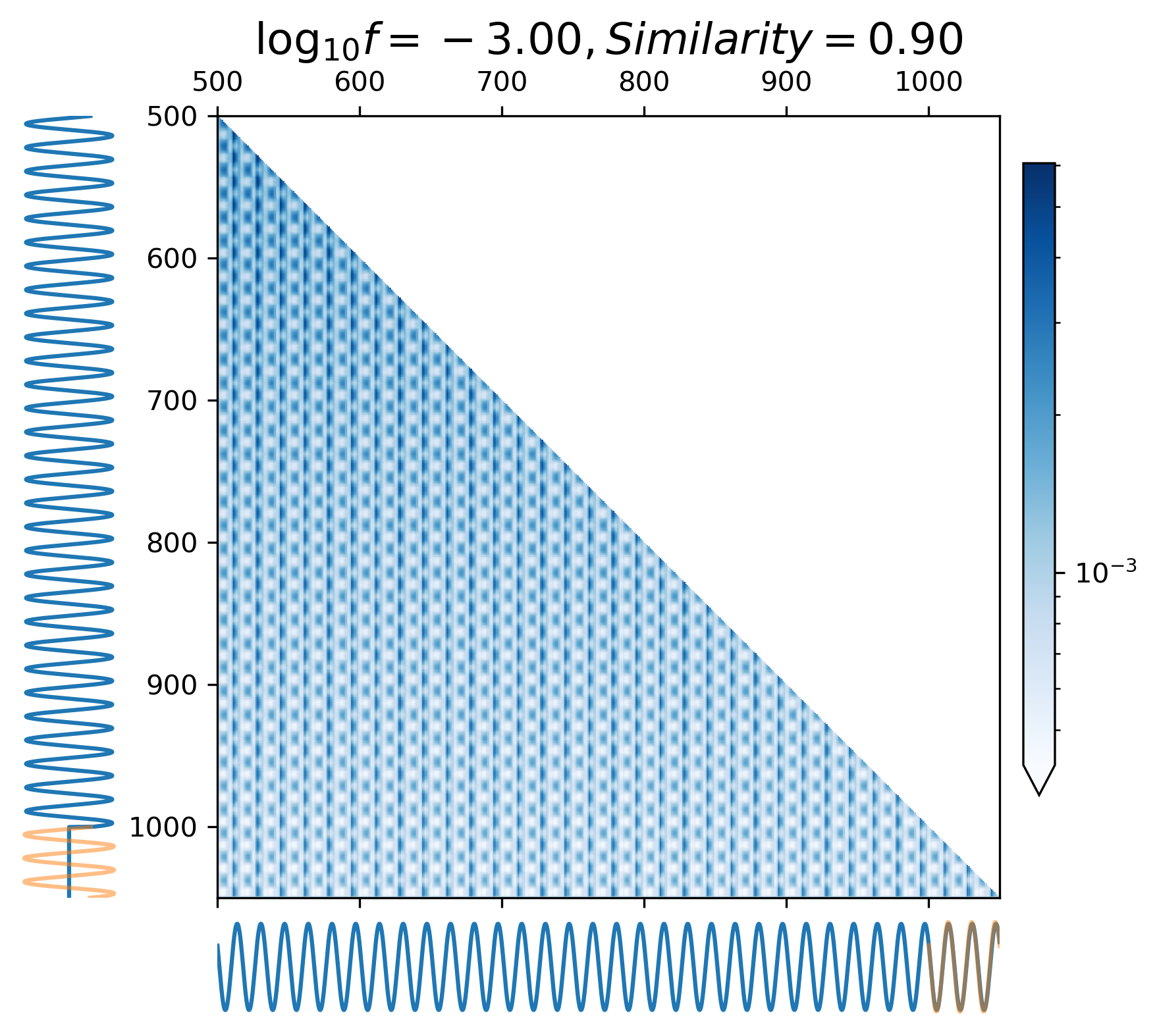}}
        \subfigure[GB 20:1 extrapolation: $f=10^{-3.5}$Hz]{\label{fig:interprete_h}
	\includegraphics[width=0.3\linewidth]{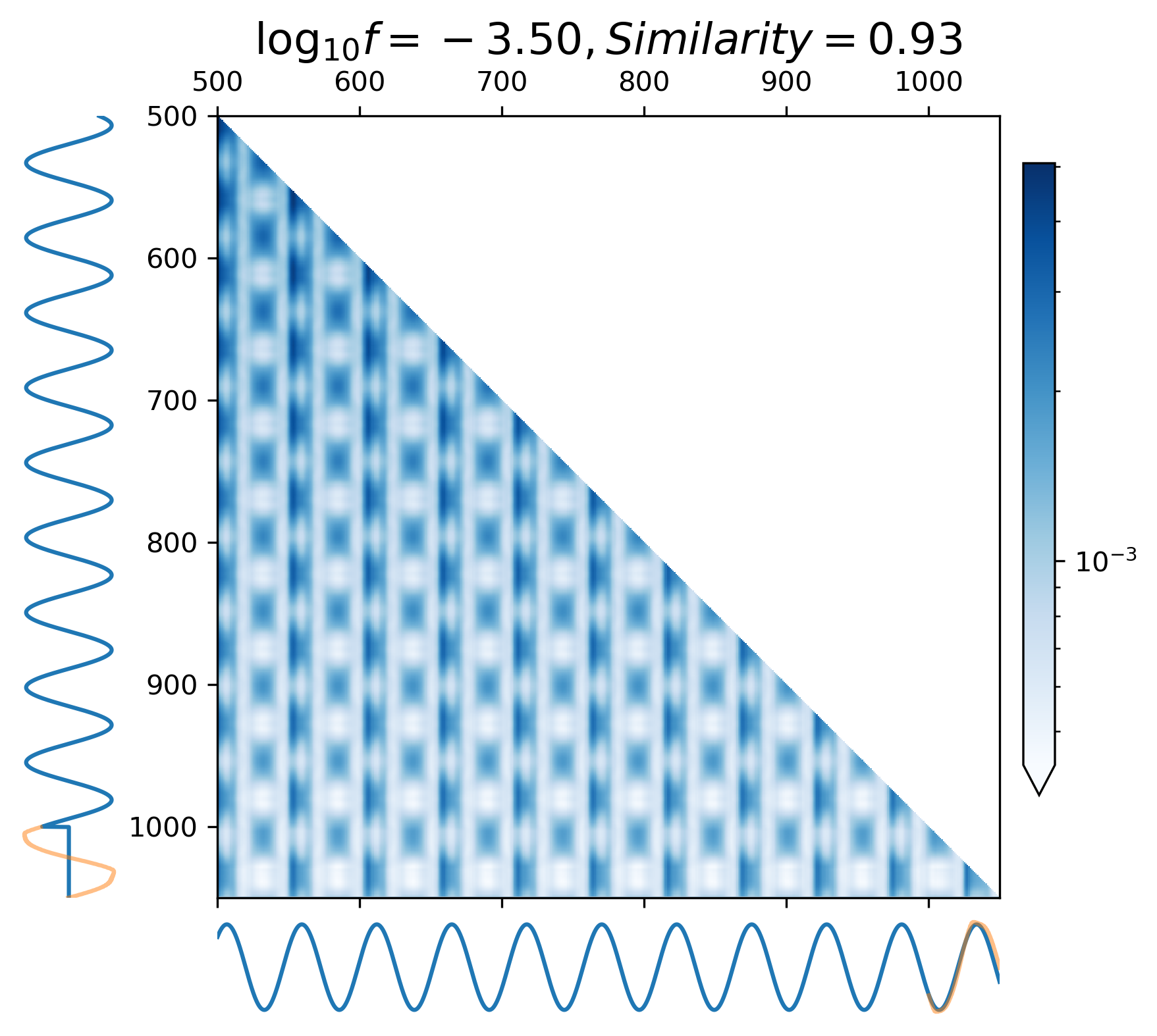}}
        \subfigure[GB 1:1 extrapolation: $f=10^{-3.52}$]{\label{fig:interprete_i}
        \includegraphics[width=0.3\linewidth]{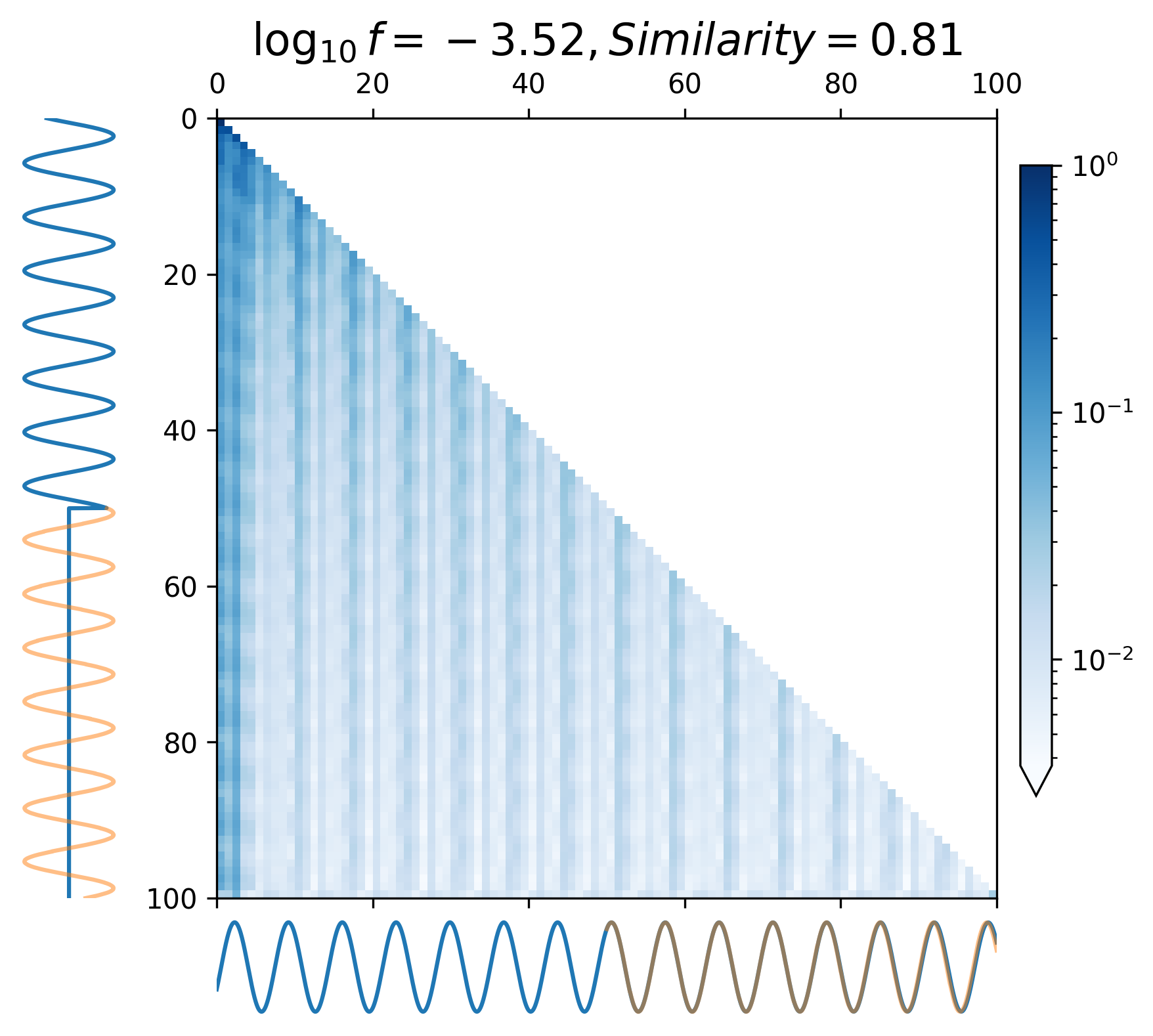}}
        \caption{\textbf{Attention maps of the last encoder layer.} For a clear presentation, only part of the attention map is displayed. The blue lines on the left and bottom panels represent the input waveforms, whose 1,001-st (or 101-st) to 1,050-th (or 50-th) tokens are padded with zero value during inference, and the orange line represents the waveform predicted by CBS-GPT. The term 'Similarity' in the title of each figure denotes the \textbf{correlation coefficient} between the waveform and the attention map. 
        }
        \label{fig:interprete}
    \end{figure*}

    The attention map (Eq. \ref{eq:attnmap}) allows us to understand the extrapolation process and attention mechanism while forecasting waveforms, making it easier to gain insight into how CBS-GPT interpret GW data. 
	\begin{equation}
		A=\frac1H\sum_{i=1}^H\text{softmax}(\frac{Q_{ji}K_{ji}^T \cdot \mathrm{mask}}{\sqrt{d}}) \:,
    \label{eq:attnmap}
	\end{equation}
	where $H$ represents all attention heads of the last encoder block.
	In Figure \ref{fig:interprete}, the vertical axis represents the model input waveform, and the horizontal axis represents the predicted waveform.
	
	When predicting continuous gravitational waveforms (EMRIs and GB), 
	the attention maps (Figure \ref{fig:interprete_d} - \ref{fig:interprete_i}) exhibit grid-like patterns that are closely related to the phase of the waveforms, with the scale of the grid expanding as the frequency decreases. 
In order to measure the similarity between the attention map and the input waveform, we introduce the \textbf{correlation coefficient} (with details described in Appendix \ref{appendix:corr_attention}).
	Overall, the average correlation coefficient of continuous waveform exceeds 0.8, which demonstrates that the model can accurately match the waveform's frequency and phase information. 
	This mode assists CBS-GPT in successfully extrapolating waveforms.
	
	As showcased in Figure \ref{fig:interprete_a} - \ref{fig:interprete_c}, during the prediction of the merge-ringdown phase of MBHB waveforms, attention primarily focuses on near-diagonal elements. 
	In contrast to continuous GW signals, the amplitude of MBHB reaches zero after the merge-ringdown, and the main focus of attention mechanism lies in the merging stage and the stage after the merge, with relatively less attention payed to inspiral phase. 

    \subsection{Potential Applications}\label{subsec:apps}
        \textbf{Complex waveforms generation.} Currently, waveform generation for high mass ratio binary black holes remains a challenging problem because of high computational cost.
        Our approach can partially alleviate this problem since CBS-GPT that trained on low mass ratio waveforms with relatively low computational cost can be applied to high mass ratio waveform generation. 
        This generalization characteristic, as shown in Figure \ref{fig:showcase_gen}, demonstrates that the model can learn intrinsic features and can be applied to waveform extrapolation of a broader parameter space.
        By incorporating simulations based on numerical relativity, we may build a waveform template bank by extrapolating more complex and computation-intensive waveforms. 
        For burst wave sources such as MBHB, waveform generation time of CBS-GPT for a single waveform is less than 100ms on a single NVIDIA V100 GPU.
        With the rapid development of GPU computing power, CBS-GPT presents the potential for high-speed template waveform generation.

        \textbf{Gap imputation.}       
        In space-based GW detectors, the presence of data gaps due to data transmission, satellite attitude adjustments, and unidentified glitches can significantly impact the precision of waveform parameter estimation.
        Our waveform extrapolation method is promising to accomplish the task of waveform imputation, and by integrating with successive denoising models \cite{torres-forneDenoisingGravitationalWave2016, weiGravitationalWaveDenoising2020, akhshiTemplatefreeApproachWaveform2021, Ren2022, zhao_space-based_2023, weiGravitationalWaveDenoising2020, chatterjeeExtractionBinaryBlack2021a, ormistonNoiseReductionGravitationalwave2020b}, parameter estimation accuracy can be further enhanced \cite{deyEffectDataGaps2021a}.
        
        \textbf{Model Design Guidance.} We established a more convenient method for visualizing and quantifying attention maps, offering guidance for transformer-based models design in the GW research realm.  
        Our results also demonstrate that attention mechanism can be leveraged to establish more robust deep learning models that are specifically tailored for GW astronomy.
        
\section{Conclusion}
    \label{sec:con}
	In this paper, we introduce the CBS-GPT model, consisting of hybrid embedding and encoder blocks.
	The CBS-GPT is applied to predict GW waveforms after the TDI 2.0 response.
        We investigated two scenarios of different extrapolation ratios between input and predicted waveform length.
        Different models are trained for MBHB, EMRIs, and GB.
        In the 20:1 and 1:1 extrapolation scenarios, the average overlaps between the predicted waveform and the target waveform of MBHB, EMRIs, and GB reach 0.981, 0.912, 0.991, and 0.990, 0.807, 0.991, respectively. 
        EMRIs exhibited poorer performance in the 1:1 extrapolation due to their complex waveform patterns and rich amplitude variations caused by eccentricity. 
        We also proved the strong generalization of CBS-GPT on MBHB waveforms.
 
	Moreover, we introduced a correlation coefficient and found that the correlation between hidden parameters of CBS-GPT and waveform was relatively high, which indicated that the model could learn waveform's phase information extremely well.
	Overall, our results show that CBS-GPT has the ability to comprehend detailed waveform properties and make predictions over varied frequencies. 
 We are confident that in the future, large AI models such as CBS-GPT can be applied to GW data processing tasks including complex waveforms generation and gap imputation.
	
	\begin{acknowledgments}
		This research was supported by the Peng Cheng Laboratory and Peng Cheng Cloud-Brain. 
		This work was also supported in part by the National Key Research and Development Program of China Grant No. 2021YFC2203001 and in part by the NSFC (No. 11920101003 and No. 12021003). 
		Z.C. was supported by the “Interdisciplinary Research Funds of Beijing Normal University" and CAS Project for Young Scientists in Basic Research YSBR-006.
	\end{acknowledgments}
	
	\appendix
	\section{Correlation coefficient between waveform and hidden parameters}
	\label{appendix:corr_attention}
        To evaluate the correlation between the attention map's grid-like pattern and the waveform, we introduce the \textbf{correlation coefficient} between the waveform and hidden parameters (or attention map). This coefficient assesses the level of correlation and demonstrates the attention map's ability to capture phase information.
	Firstly, we compute the mean value of each token of the patched waveform to get the sequence $M$.
	Subsequently, the outer product of $M$ is computed, resulting in the auto-correlation matrix. 
	As the attention map $A$ (Eq. \ref{eq:attnmap}) is processed by masking and normalization, we do a similar adjustment to the auto-correlation matrix:
	\begin{equation}
		\begin{aligned}
			&R_{\text{mask}} = \text{Mask}(M \otimes M - \text{min}(M \otimes M)), \\
			&R_{\text{Norm}} = \text{RowNorm}\left(
			R_{\text{mask}}
			\right), 
		\end{aligned}
		\label{eq:norm}
	\end{equation}
	where $\text{RowNorm}(\cdot)$ denotes the normalization of each row of the matrix and $\text{Mask}(\cdot)$ is consistent with the mask method of Section \ref{subsec:tni}.
 
    To assess the correlation between the two matrices, we calculate the Pearson correlation coefficient between the flattened attention map $A$ and flattened $R_{\text{Norm}}$: 
	\begin{widetext}
		\begin{equation}
			\begin{aligned}
				\rho_{A, R_{\text{Norm}}} 
				&= \rho \left\{\text{Flatten}(A), 
				\text{Flatten}(R_{\text{Norm}}) \right\}\\
				&= \frac{n\sum_1^n {A_{F}}_i {R_{F}}_i- \sum_1^n {A_{F}}_i \sum_1^n {R_{F}}_i}
				{\sqrt{n \sum_1^n {A_{F}}_i^2-(\sum_1^n {A_{F}}_i)^2}
					\sqrt{n\sum_1^n {R_{F}}_i^2- (\sum_1^n {R_{F}}_i)^2}}
			\end{aligned}
			\label{eq:pearson}
		\end{equation}
	\end{widetext}
	where $\text{Flatten}(\cdot)$ denotes flattening the matrix into one dimension, $A_F$ and $R_F$ represent the flattened vector of $A$ and $R_{\text{Norm}}$ respectively, and $n$ represents the length after flattening. 
	Finally, $\rho_{A, R_{\text{Norm}}}$ is defined as the correlation coefficient between waveform and hidden parameters.
	
	\bibliographystyle{myREVTeX4-2}
	\bibliography{references}
\end{document}